\documentclass[12pt]{article}

\pdfoutput=1
\usepackage{color}
\usepackage{epsfig, palatino}
\usepackage{pstricks,pst-node,pst-tree}
\usepackage{epic}
\usepackage{mathrsfs}
\usepackage{ae} 
\usepackage[T1]{fontenc}
\usepackage[ansinew]{inputenc}
\usepackage{amsmath}
\usepackage{amssymb}
\usepackage{comment}
\usepackage{graphicx}
\usepackage{ulem}
\usepackage{xcolor}
\definecolor{darkblue}{cmyk}{0.9,0.9,0,0}
\definecolor{c1}{RGB}{236, 0, 11}
\definecolor{c2}{RGB}{0, 163, 244}
\definecolor{c3}{RGB}{170, 4, 212}
\definecolor{c4}{RGB}{255, 153, 0}
\definecolor{c5}{RGB}{45, 224, 0}
\usepackage[colorlinks=true,linkcolor=darkblue,citecolor=darkblue,urlcolor=darkblue]{hyperref}
\usepackage{cite}
\usepackage{hyperref}
\usepackage{wasysym}
\usepackage{varioref}
\usepackage{makeidx}
\usepackage[english]{babel}
\usepackage{simplewick}
\usepackage{array}
\usepackage{multirow}

\usepackage[font={small}]{caption}

\newcommand{\beq}{\begin{equation}}
\newcommand{\eeq}{\end{equation}}
\newcommand{\beqq}{\begin{equation*}}
\newcommand{\eeqq}{\end{equation*}}
\newcommand\beqa{\begin{eqnarray}}
\newcommand\eeqa{\end{eqnarray}}
\newcommand\beqaa{\begin{eqnarray*}}
\newcommand\eeqaa{\end{eqnarray*}}
\newcommand\bea{\begin{array}}
\newcommand\eea{\end{array}}

\newcommand{\neqa}{\nonumber\end{eqnarray}}

\renewcommand{\d}{\partial}

\newcommand{\<}{{\langle}}
\renewcommand{\>}{{\rangle}}

\newcommand{\cL}{{\cal L}}

\newcommand{\re}{\relax{\rm I\kern-.18em R}}

\renewcommand{\sp}{p\hspace{-.40em}/}

\definecolor{darkgreen}{rgb}{0.0, 0.45, 0.0}

\def\XXint#1#2#3{{\setbox0=\hbox{$#1{#2#3}{\int}$}
\vcenter{\hbox{$#2#3$}}\kern-.5\wd0}}

\def\su2{{SU(2)}}

\def\[{\left[}
\def\]{\right]}

\def\({\left(}
\def\){\right)}
\def\[{\left[}
\def\]{\right]}

\def\<{\langle}
\def\>{\rangle}

\def\i2{\frac{i}{2}}

\def\spi{\relax{\rm \pi\kern-0.5em /}}
\def\sA{\relax{\rm A\kern-0.5em /}}
\def\sp{\relax{\rm p\kern-0.5em /}}
\def\sd{\relax{\rm \d\kern-0.5em /}}
\def\sk{\relax{\rm k\kern-0.5em /}}
\def\sn{\relax{\rm n\kern-0.5em /}}
\def\sl{\relax{\rm l\kern-0.5em /}}
\def\sP{\relax{\rm P\kern-0.7em /}}
\def\sBethe{\relax{\rm \Bethe\kern-0.5em /}}

\def\cD{{\cal D}}

\def\cS{{\cal S}}

\def\2F1{\,_2{\rm F}_1}

\makeindex

\begin{document}

\thispagestyle{empty}

\renewcommand{\thefootnote}{\fnsymbol{footnote}}
\setcounter{page}{1}
\setcounter{footnote}{0}
\setcounter{figure}{0}

\begin{flushright}
\vspace{-1cm}
\small{DESY-24-004}
\vspace{1cm}
\end{flushright}

\begin{center}
$$$$
{\Large\textbf{\mathversion{bold}
Two loop five point integrals: light, heavy and large spin correlators
}\par}

\vspace{1.0cm}

\textrm{Carlos Bercini$^\text{\tiny 1}$, Bruno Fernandes$^\text{\tiny 2}$,Vasco Gon\c{c}alves$^\text{\tiny 2}$  }
\\ \vspace{1.2cm}
\footnotesize{\textit{
$^\text{\tiny 1}$ Deutsches Elektronen-Synchrotron DESY, Notkestr. 85, 22607 Hamburg, Germany\\
$^\text{\tiny 2}$Centro de F$\acute{\imath}$sica do Porto, Departamento de F$\acute{\imath}$sica e Astronomia,
Faculdade de Ci$\hat{e}$ncias da Universidade do Porto, Rua do Campo Alegre 687, 4169-007 Porto, Portugal\\
}  
\vspace{4mm}
}

\par\vspace{1.5cm}

\textbf{Abstract}\vspace{2mm}
\end{center}
We evaluated all two loop conformal integrals appearing in five point correlation functions of protected operators of $\mathcal{N}=4$ Super Yang-Mills in several kinematical regimes. Starting from the correlation function of the lightest operators of the theory, we were able to extract structure constants of up to two spinning operators for small and large values of polarizations and spin. We conjectured an universal all loop behaviour for the large spin small polarization structure constants and comment on the subtleties of analytically continuing it from finite to large spin. We also consider correlation functions of heavier operators that get factorized in the more fundamental object called decagon. We fixed this object at two loops  in general kinematics and studied its physical properties under OPE and null limits.

\noindent

\setcounter{page}{1}
\renewcommand{\thefootnote}{\arabic{footnote}}
\setcounter{footnote}{0}

\setcounter{tocdepth}{2}

 \def\nref#1{{(\ref{#1})}}

\newpage

\tableofcontents

\parskip 5pt plus 1pt   \jot = 1.5ex

\section{Introduction} 
Correlation functions of local operators are one of the most natural observables to consider in a conformal field theory.  One of the most famous and interesting conformal field theories is the four dimensional $\mathcal{N}=4$ super Yang Mills that is dual, through the AdS/CFT correspondence, to type IIB super string theory in $AdS_5\times S_5$. 

An important class of such operators in this theory are the scalar single-trace half-BPS operators. Even though supersymmetry prevents their two and three point functions to receive quantum
corrections \cite{Intriligator:1998ig}, their higher point correlation functions encode a plethora of information about interesting physical observables of the theory via a web of dualities between correlation functions, null polygon Wilson loops and gluon scattering amplitudes \cite{Alday:2010zy,Eden:2010zz,Alday:2007hr,Brandhuber:2007yx,Drummond:2007cf,Alday:2010ku}. Their correlation functions can be written as
\begin{align}
\langle \mathcal{O}_{k_1} \mathcal{O}_{k_2} \dots \mathcal{O}_{k_n}\rangle=
\sum_{\ell=0}^\infty \lambda^{\ell} \mathcal{G}_n^{(\ell)}
\,, \quad \quad {\rm{with}} \quad \quad
\lambda = \frac{g_{YM}^2N_c}{4 \pi^2} \, .
\label{eq:Gnloop}
\end{align}
where the half-BPS operators $\mathcal{O}_{k_i}= \text{Tr}(Y_i\cdot\Phi(x_i) )^{k_i}$ are defined with the help of an auxiliary six-dimensional null vector $Y_i$ and $N_c$ is the number of colors of the SU$(N_c)$ gauge group, which will always taken to be large.

The first non-trivial higher point correlators are the four point functions, which have been studied since the early days of AdS/CFT correspondence. Their integrands have been computed perturbatively in the planar limit up to ten loops in the t'Hooft coupling $\lambda$, see \cite{Eden:2012tu,Chicherin:2015edu, Bourjaily:2016evz,Chicherin:2018avq, Caron-Huot:2021usw}. Meanwhile, for higher point functions, beyond the one loop result of \cite{Drukker:2008pi}, very few cases have been worked out \cite{Goncalves:2019znr,Bargheer:2018jvq,Fleury:2020ykw,Goncalves:2023oyx}.

The central object we will consider in this paper is one of these exceptions: the integrand of the five point correlation function of the lowest weight protected operator, which has been recently computed up to three loops in \cite{Bargheer:2022sfd}. At two loops the five point correlator can be expressed in terms of several new conformal integrals 

\begin{align}
&\mathcal{L}_{\textcolor{c1}{1};\textcolor{c2}{2}\textcolor{c3}{3};\textcolor{c4}{4}\textcolor{c5}{5}}=\int \frac{d^4x_7d^4x_8}{x_{\textcolor{c1}{1}7}^2x_{\textcolor{c1}{1}8}^2  x_{\textcolor{c2}{2}7}^2x_{\textcolor{c3}{3}7}^2x_{\textcolor{c4}{4}8}^2x_{\textcolor{c5}{5}8}^2x_{78}^2} 
\label{intL}\\
&\mathcal{S}_{\textcolor{c1}{1};\textcolor{c2}{2}\textcolor{c3}{3};\textcolor{c4}{4}\textcolor{c5}{5}} = \int \frac{d^4x_7\, d^4x_8 x_{\textcolor{c1}{1}7}^2}{x_{\textcolor{c1}{1}8}^2 x_{\textcolor{c2}{2}7}^2x_{\textcolor{c3}{3}7}^2x_{\textcolor{c3}{3}8}^2 x_{\textcolor{c4}{4}7}^2 x_{\textcolor{c5}{5}8}^2 x_{\textcolor{c5}{5}7}^2x_{78}^2}
\label{intS}\\
&D_{\textcolor{c2}{2}\textcolor{c4}{4},\textcolor{c1}{1}\textcolor{c3}{3}\textcolor{c5}{5}}= \int \frac{ d^4x_7d^4x_8 x_{\textcolor{c2}{2}7}^2x_{\textcolor{c4}{4}8}^2 }{x_{\textcolor{c1}{1}7}^2x_{\textcolor{c1}{1}8}^2 x_{\textcolor{c2}{2}8}^2 x_{\textcolor{c3}{3}7}^2x_{\textcolor{c3}{3}8}^2 x_{\textcolor{c4}{4}7}^2 x_{\textcolor{c5}{5}7}^2x_{\textcolor{c5}{5}8}^2x_{78}^2 } \,.
\label{intD}
\end{align}

Contrary to their four point two loop counterparts, these integrals are not known in closed form. In this paper we managed to compute these integrals for particular kinematic configurations: the plane kinematics (all points in plane), the light-cone limit (points becoming null separated) and the OPE limit (points approaching each other). This allowed us to evaluate the five point correlator and study its physical properties under these several regimes.

During the last couple of years there has been increasing interest in analysing the constraints coming from the conformal bootstrap equations for more than four operators. These are equations that involve kinematical functions known as conformal blocks and products of OPE coefficients \cite{Poland:2023vpn,Poland:2023bny}. Solving these equations in full generality, just like evaluating conformal integrals, is a hard business and one typically restricts to kinematical limits and uses certain assumptions. A common central assumption is that OPE coefficients of spinning operators are reasonably well behaved in the large spin limit, being given by analytical functions that depend logarithmically on the spins. For instance, in \cite{Bercini:2020msp} using this assumption together with consistency of the conformal bootstrap equations lead to fixing the form of a five point correlation function when neighbouring distances are taken to be null as well as fixing the universal behaviour of these large spin OPE coefficients.

One possible way to probe the correctness of these assumptions is by going through an explicit computation in a specific model. One of the main goals of this paper is to do precisely that, for the five point correlator of $20^\prime$ operators in $\mathcal{N}=4$ SYM. Using the results evaluated here for the integrals (\ref{intL},\ref{intS},\ref{intD}) we can compute this correlator in several kinematics and extract OPE coefficients of two spinning operators for arbitrary values of spins and polarizations.

For finite values of spin, we can compare our results with \cite{Bianchi:2019jpy,Bianchi:2021ohn} where some of these structure constants were computed before. For large values of spin we can make contact with the analytical bootstrap results of \cite{Bercini:2020msp}. However, we observed an important feature that first presents itself at two loops: \textit{the large spin limit of the finite spin result is different than the expected large spin behaviour} and consequently in contradiction with the assumptions that are usually considered in large spin analysis \cite{Alday:2013cwa,Bercini:2021jti,Antunes:2021kmm,Kaviraj:2022wbw}. Due to our analytical control over the correlator, we were able to clean this apparent paradox.

Another interesting type of correlation functions are the so-called simplest correlators, introduced in \cite{Coronado:2018cxj}. They consist of operators with large values of $R$-charge ($k_i \gg 1$) that factorize the correlator as the square of a simpler object. For four point functions this object is called \textit{octagon} ($\mathbb{O}$) which was bootstrapped in \cite{Coronado:2018cxj} at all orders in the t'Hooft coupling. 

The five point incarnation of such a factorized object is the \textit{decagon} ($\mathbb{D}$), which was computed via integrability at two loops in the plane kinematics \cite{Fleury:2020ykw}. By demanding that this ansatz is consistent with the Lagragian insertion method \cite{Eden:2012tu,Bargheer:2022sfd} at two loops we were able to lift the decagon away from the plane into general kinematics. This is one of our main results and it is presented in (\ref{decResult}).
	
Furthermore, since the decagon is given in terms of the conformal integrals (\ref{intL},\ref{intS}), we can now study integrated properties of this object. More precisely, we checked that the decagon satisfies the same properties of the octagon: Steinmann relations and boundary data constraints such as the stampedes \cite{Olivucci:2021pss,Olivucci:2022aza}, the central difference between these objects being the basis of functions, which is far more complicated in the five point function case.

An important application of our results is the computation of these two correlators in the limit where the five points approach the cusps of a null polygon. We can explicitly check that the leading divergences of these correlators are governed by two different quantities \cite{Caron-Huot:2021usw,Bork:2022vat}, $\Gamma_{\text{cusp}}$ for the $20^\prime$ correlator and $\Gamma_{\text{oct}}$ for the decagon, as presented in equations \eqref{nullC} and \eqref{decNull} respectively. Furthermore, for the $20^\prime$ correlator we also found a perfect agreement with the bootstrapped results of \cite{Bercini:2020msp}, hence a match with the null pentagon Wilson loop.

This paper is divided into two main parts. In part \ref{ints} we evaluate the two loop integrals (\ref{intL},\ref{intS},\ref{intD}) in the several kinematics limits aforementioned. In part \ref{corr} we study the five point function of $20^\prime$ operators and the decagon in these regimes.

\part{Integrals}
\label{ints}
This first part is technical and details how we computed the complicated conformal integrals (\ref{intL},\ref{intS},\ref{intD}). Some aspects of our computation can be, in principle, applied in more general cases, being either higher loops or higher number of points. This entire part can be skipped by readers that are not concerned with the integrals themselves but rather with the linear combination of integrals known as correlators.

\section{Conformal Integrals}
Five point correlators are functions that depend on five cross-ratios
\begin{equation}
u_1=\frac{x_{12}^2x_{35}^2}{x_{13}^2x_{25}^2}\,, \qquad  u_{i+1}=u_i\big|_{x_{j}\rightarrow x_{j+1} } \,. \label{eq:conformalCrossratios}
\end{equation}
Ideally one would like to have these integrals expressed in terms of special functions (for example, multiple polylogarithms \cite{Goncharov:1998kja,Gehrmann:2001jv}) of the cross-ratios but in practice this is a hard task. However, there are special configurations where the expressions of these integrals simplify. One such example is when we set all the five points to lie on a common plane, see \cite{Fleury:2020ykw,Aprile:2023gnh}. Another example that is of interest to this paper is the limit where some points are approaching the light-cone of others. The goal of this section is to explain how we evaluate these integrals in these two kinematics.  We will use two different methods to compute them: the plane configuration will be evaluated by rewriting the integrals using the Schwinger parametrization, while the light-cone limit will be evaluated by dividing the integrals into simpler regions, which are further simplified by identities following from integration by parts identities.

Before diving into the computations, we establish some notation. All the two loop conformal integrals considered in this paper are of the form
\begin{align}
I = \int \frac{d^dx_7d^dx_8}{x_{78}^2\prod_{i=1}^5\prod_{j=7}^8(x_{ij}^2)^{a_{ij}}  }\label{eq:conformalintegralI}
\end{align}
with the constants $a_{ij}$ being integers subject to the conformal constraint $\sum_i a_{ij}=d-1$. Lastly, we recall the names that these integrals usually undergo: $\mathcal{L}$ integrals are the \textit{double-box} \eqref{intL}, $\mathcal{S}$ integrals are the \textit{penta-box} \eqref{intS} and $\mathcal{D}$ integrals are the \textit{double-penta} \eqref{intD}.

\subsection{Parametric integration}
Schwinger parametrization is based on the following simple identity
\begin{align}
\frac{1}{A^n} =\frac{1}{\Gamma(n)}\int_{0}^{\infty} d\alpha \alpha^{n-1} e^{-\alpha\, A}
\label{scheq}
\end{align}
and can be used to derive a representation of the integrals 
\begin{align}
I(a_{ij})= \left(\prod_{i}\int_{0}^{\infty}\alpha_{i}^{a_i-1}d\alpha_i  \right) \frac{\mathcal{U}^{\omega-\frac{d}{2}} }{\mathcal{F}^{\omega}}\delta(1-\sum_i\alpha_i),  \ \ \ \omega=1+\sum_{i=1}^5\sum_{j=7}^8a_{ij} -d
\end{align}
where $\mathcal{U}, \mathcal{F}$ are graph polynomials (see \cite{Golz:2015rea} for more details) and $i$ runs over the propapagators appearing in (\ref{eq:conformalintegralI}) while the $a_i$ are their powers.

One of the advantages of this representation is that the space time integration of the original integral is already done. Given that the integral (\ref{eq:conformalintegralI}) is conformal it is useful to send one \textit{complicated} external point to infinity resulting in a simpler parametrization. A natural choice is the point that appears the most times in the integration or the point that appears as a numerator. When the integral has no numerator or just one numerator like \eqref{intL} and \eqref{intS} this procedure can be easily done. However, for integrals with more than one numerator like \eqref{intD} one needs to change the expression \eqref{scheq} to be valid\footnote{Note that in \eqref{scheq} negative $n$ values correspond to poles in the gamma function.} also for integrals with numerators \cite{Golz:2015rea}.

After sending the \textit{complicated} points to infinity the the integrals $\cL$ and $\cS$ are ready to be evaluated, provided that we treat carefully the effects of sending this point to infinity. We were only able to do this by choosing a configuration where the points $x_i$ lie on a common plane making the integrals $\cL$ and $\cS$ become linear reducible \cite{Brown:2008um}. In the end the integration over the parametric variables $\alpha_i$ can be accomplished with use of a computer package \texttt{HyperInt} developed in \cite{Panzer:2014caa}. The result is expressed in terms of multiple polylogarithms multiplied by rational functions of cross-ratios, which we attach in an auxiliary \texttt{Mathematica} file.

The $\cL$-type integrals have a factorized cross-ratio dependence prefactor, for example
\begin{align}
\mathcal{L}_{5,12,34} &= \frac{1}{z\bar{y}-y\bar{z}}\Bigg( G(0;y)G(0;\bar{z})G(1;\bar{y})G(1-y;z) + G\left(\frac{y}{1-z};\bar{y}\right)G\left(1-y,1,0;z\right)+\nonumber \\
&G\left(\frac{y-(1-z)\bar{y}}{y},\frac{y(1-\bar{y})z}{yz + \bar{y}(1-y-z)},1-\bar{y},1;\bar{z}\right)+\dots\Bigg)
\end{align}
where the $\dots$ stand for other multiple polylogarithms recalled in appendix \ref{appInts}, which contain the following list of letters 
\begin{align}
\{y,z,\bar{y},\bar{z},1-y,1-z,1-\bar{y},1-\bar{z},y-\bar{y},1-y-z,1-\bar{y}-\bar{z},z-\bar{z},&\nonumber\\
^,y-\bar{y}(1-z)-y\bar{z},z-\bar{z}(1-y)-\bar{y}z,y\bar{y}\bar{z}-\bar{y}\bar{z}(1-z)+yz(1-\bar{y}-\bar{z}\}
\end{align}
and the cross-ratios in the plane kinematics can be related to the original $u_i$ of (\ref{eq:conformalCrossratios}) via
\begin{align}
u_1= z \bar{z},u_2&= \frac{(1-y) (1-\bar{y} ) (z-1) (\bar{z}-1)}{(y+z-1) (\bar{y} +\bar{z}-1)},u_3= \frac{y \bar{y}   }{(y+z-1) (\bar{y} +\bar{z}-1)},\nonumber\\
&u_4= \frac{1}{(1-y) (1-\bar{y} )},u_5= \frac{(y+z-1) (\bar{y} +\bar{z}-1)}{(1-y) (1-\bar{y}  )}\,.
\label{eq:planecrossratios}
\end{align}

These $\mathcal{L}$ integrals are by far the simplest and their overall prefactor can be independently computed by doing an analysis of the leading singularity of the integral, see \cite{Drummond:2013nda}. On the other hand, the $\cS$ integrals do not have a common prefactor and one cannot do such a simple analysis for a sanity check. Due to the presence of numerators, the $\cD$  integrals are more difficult to compute using this method, however all of these integrals can be computed using the other methods explained below.

The plane kinematics is useful to obtain explicit expressions for the two loop five point conformal integrals, but it is not enough to study some interesting physical limits such as the OPE or sequential light-like limits. However, it is possible to obtain corrections around plane kinematics by noticing that both $\cL$ and $\cS$ integrals satisfy differential equations coming from the four-dimensional identity
\begin{align}
\nabla_{i} \frac{1}{x_{i7}^2} = -4\pi^2\delta(x_{i7}).
\end{align}

We can only apply this simple relation to integrals where the point $x_i$ appears once in the propagator, there are four of these equations for each $\cL$ integral, two for each $S$ integral and none for the $\mathcal{D}$ integrals. As explained in \cite{Fleury:2020ykw} (see also \cite{soton428038}), the differential equation can be  expressed  in terms of the cross-ratios (\ref{eq:planecrossratios}) plus a fifth cross-ratio, $\eta$, that controls the out of the plane kinematics, defined by
\begin{align}
u_2= \frac{(1-y) (1-\bar{y} ) (z-1) (\bar{z}-1)}{(y+z-1) (\bar{y} +\bar{z}-1)}(1+\eta).
\end{align}
It is easy to see that when $\eta=0$  one recovers the plane kinematics \eqref{eq:planecrossratios}. The differential equation, when written in these variables, has some terms that interlink the plane and out-of-plane kinematics \cite{Fleury:2020ykw}, which can be used to generate a series expansion in $\eta$ for the integrals. 

Thus, one can start with the plane expression for the integrals and, in principle, use the differential equations to generate the full dependence in the cross-ratios\footnote{Apart from the integrals $\mathcal{D}$, which we could not easily compute nor lift out of the plane}. In practice this method is very inefficient and we managed to compute only the first few orders in the $\eta$ expansion. Nonetheless this gave us very non-trivial expressions for the integrals, which were used to verify the correctness of our computations when using the different methods described below.

\subsection{Asymptotic expansions}
Instead of sending all external points to the plane, we will compute the integrals in a kinematic configuration that makes contact with more physical regimes of the correlator: the single light-cone limit, where points $12$ and $34$ become null separated of each other. We will review the method of asymptotic expansions and explain how it can be applied efficiently to obtain our two loop integrals in the limit $x_{12}^2,x_{34}^2\rightarrow 0$, or in term of cross-ratios when $u_1,u_3 \to 0$. Given that this method has been applied extensively in integrals similar to the ones we considered here, we will just give a lightning review and refer to following references for more details \cite{Smirnov:2002pj,Eden:2012rr,Goncalves:2016vir,Georgoudis:2017meq,Bargheer:2022sfd}. 

Whenever we have a parameter, like these null distances, which is sent to zero we can apply the asymptotic expansion method. The main idea behind the method is to divide each integration variable into two regions: one where the integration variable is small enough to be comparable to the small parameter and other where it is large enough such that the parameter can be ignored. We will take the limits sequentially, {\it i.e.} first we take $x_{12}^2\rightarrow 0$ and only then we take $x_{34}^2\rightarrow 0$. In general, for an $l$ loop integral there are $2^{l}$ regions when we take a distance to be light-like, in particular, for the two loop case we have the following four regions
\begin{equation}
1.\,\, x_7 \ll 1\,, x_8\ll 1\,; \quad 2.\,\, x_7 \ll 1\,, x_8\gg 1\,; \quad 3.\,\, x_7 \gg 1\,, x_8\ll 1\,;\quad 4.\,\, x_7 \gg 1\,, x_8\gg 1\,.
\label{4regions}
\end{equation}

The advantage of doing this separation is that we are allowed to simplify the integrand accordingly in each region by using the identities
\begin{align}
	\label{eq:asymptoticExpansion}
\frac{1}{(x_{ij}^2)^b} = \frac{1}{(x_{1j}^2)^b}\sum_{n_b=0}^{\infty} {{-b}\choose{n_{b}} }\frac{(x_{1i}^2-2x_{1i}\cdot x_{1j})^{n_b}}{(x_{1j}^2)^{n_b}}
\end{align}
where we assumed $x_j$ to be large and $x_i$ to be small. Consider for example, the integral $\mathcal{L}_{5,13,24}$ and let us apply the prescription above 
\begin{align}
&\hspace{-1cm}\mathcal{L}_{5,13,24}=\int \frac{d^4x_7d^4x_8}{x_{17}^2x_{37}^2x_{28}^2x_{48}^2x_{78}^2} = I_1+I_2+I_3+I_4,\\
I_1=\sum_{n,m=0}^{\infty}&\int \frac{d^dx_7d^dx_8\,(2x_{13}\cdot x_{37}-x_{17}^2)^{n}  (2x_{14}\cdot x_{48}-x_{18}^2)^{m}  }{(x_{13}^2)^{1+n}(x_{14}^2)^{1+m}x_{17}^2x_{18}^2x_{28}^2x_{78}^2}\nonumber,\\
I_3= \sum_{n,m=0}^{\infty}&\int \frac{d^dx_7d^dx_8\,(2x_{14}\cdot x_{48}-x_{18}^2)^{m} \,(2x_{17}\cdot x_{18}-x_{18}^2)^{n}  }{ (x_{14}^2)^{1+m}x_{17}^2x_{18}^2x_{28}^2 x_{37}^2(x_{17}^2)^{1+n} },\, \\
&I_4= \sum_{n=0}^{\infty}\int \frac{d^dx_7d^dx_8 (2x_{12}\cdot x_{18}-x_{12}^2)^{n}  }{x_{17}^2(x_{18}^2)^{2+n }x_{37}^2x_{48}^2 x_{78}^2}\nonumber.
\end{align}
where we assumed $x_{12}^2\rightarrow 0$ and have taken $x_5$ to infinity. Let us note we only have written explicitly $I_3$ and not $I_2$ because the latter does not contribute - it evaluates to zero, since it is scaleless\cite{Abreu:2022mfk}. One important point is that this method allows to divide/regulate the dependence on the null distance $x_{12}^2$. For example, region 4 is effectively the original integral evaluated exactly at $x_{ 12}^2 = 0$, while the remaining regions are the ones that can contain a non-trivial dependence in $x_{12}^2$ and in particular can generate divergent terms such as $\ln x_{12}^2$. 

It should also be evident from the integrals above that each $I_i$ is simpler than the original integral. The idea is that now we can sytematically identify the dependence on the vanishing distance and drop it. Moreover, one can apply the asymptotic expansion method again to each individual region, but this time considering that the distance $x_{34}^2$ is becoming null. This will give rise to several simpler regions that we were able to evaluate. In some cases it was not even necessary to consider the $x_{34}^2\to0$ regime since the integrals became simple enough to be computed, as is the case of $I_1$ above. 

As mentioned before, the $\cD$ integrals cannot be easily computed using the method of parametric integration. For this reason, we will devote the remainder of this section to this particular type of integral. This should also serve as a paradigmatic example that features all the details of our computation and that can be applied to all other simpler integrals. Our specific example will be
\begin{equation}
 \cD_{24,135}= \int \frac{ d^dx_7d^dx_8 x_{27}^2x_{48}^2 }{x_{17}^2x_{18}^2 x_{28}^2 x_{37}^2x_{38}^2 x_{47}^2 x_{57}^2x_{58}^2x_{78}^2 }\,,
\end{equation} 
when considering the null limits $x_{12}^2,x_{34}^2 \to 0$, only three non-trivial regions contribute at leading order in the null distances, they are
\begin{align}
&\qquad \quad \quad \cD_{24,135} =I_3+I_{4,2} +I_{4,4},  \\
I_3 = \sum_{n,m=0}^{\infty}&\int \frac{d^dx_7d^dx_8\, (x_{17}^2-2x_{12}\cdot x_{17}) (x_{14}^2-2x_{14}\cdot x_{18} ) (2x_{13}\cdot x_{18})^{n} (2x_{17}\cdot x_{18})^{m} }{(x_{13}^2)^{1+n} x_{18}^2x_{28}^2  (x_{17}^2)^{2+m } x_{37}^2 x_{47}^2 }\nonumber\\
I_{4,2} =\sum_{n,m=0}^{\infty}\int &\left.\frac{d^dx_7d^dx_8 \, (x_{23}^2-2x_{32}\cdot x_{37})\, (x_{38}^2-2x_{38}\cdot x_{34}) (2x_{13}\cdot x_{37} )^{n} (2x_{38}\cdot x_{37} )^{m} }{(x_{13}^2)^{1+n} x_{18}^2 \,x_{37}^2\,x_{28}^2\,x_{47}^2\,(x_{38}^2)^{2+m}}\right|_{x_{12}^2=0}\nonumber\\
&\qquad \quad I_{4,4}  = \int\left. \frac{ d^dx_7d^dx_8 x_{27}^2x_{48}^2 }{x_{17}^2x_{18}^2 x_{28}^2 x_{37}^2x_{38}^2 x_{47}^2 x_{78}^2 }\right|_{x_{12}^2=x_{34}^2=0}  \nonumber. 
\end{align}
where the index indicates from which region they originated \eqref{4regions}. For example, $I_{4,2}$ comes from the region $4$ of the asymptotic expansion of $x_{12}^2\to0$ and region $2$ of the asymptotic expansion of $x_{34}^2\to0$ while $I_3$ comes from the region $3$ of the $x_{12}^2\to0$ expansion, and can already be evaluated without the need to take any further limit in $x_{34}^2$.

The first two regions have a similar structure and can be simplified using the same strategy. Let's focus on the first one, where the integration over the point $x_8$ can be done in $d=4-\epsilon$ regularization\footnote{The conformal integrals do not depend on the regulator $\epsilon$ and checking that this factor cancels after summing all the regions is a non trivial sanity check one always performs.}
\begin{align}
\int \frac{d^dx_8 (x_{13}\cdot x_{18})^{n}\, (x_{17}\cdot x_{78})^{m} }{x_{18}^2 x_{28}^2}  = \frac{ (x_{12}\cdot x_{13})^{n} (x_{17}\cdot x_{12})^{m}  }{ \epsilon (x_{12}^2)^{ \epsilon}} \frac{(1-\epsilon)_{ m+n } }{(2-2\epsilon)_{ m+n }}+\dots\label{eq:twoloopPWithnumer}
\end{align}
where the $\dots$ represent subleading terms in $x_{12}^2$. The integration over $x_7$ can also be simplified by the following identity
\begin{align}
\int \frac{d^dx_7\, (2x_{17}\cdot x_{12})^{n } }{ (x_{17}^2)^{1+n} x_{37}^2x_{47}^2 } = \frac{1}{n!}\left(z\cdot \frac{d}{dx_1}\right)^{n} \int \frac{d^dx_7}{x_{17}^2x_{37}^2x_{47}^2 }\bigg|_{z=x_{12}}.\label{eq:SimpleOneLoop3pt}
\end{align}
Given that (\ref{eq:SimpleOneLoop3pt}) is a convergent integral one needs to series expand  (\ref{eq:twoloopPWithnumer}) up to subleading order in $\epsilon$
\begin{align}
\frac{(1-\epsilon)_{ n} }{(2-2\epsilon)_{ n } } = \frac{1}{1+n} +\frac{\epsilon}{(1+n)^2}\left((1+n)S_{1+n}  -(2n+1)\right)\label{eq:numeratorint2ptexpression}.
\end{align}
By combining \eqref{eq:twoloopPWithnumer}, \eqref{eq:SimpleOneLoop3pt} and \eqref{eq:numeratorint2ptexpression} in the region integral $I_3$, one can integrate each term of the sum in $n$ and $m$ and write this integral as some power law expansion in the cross-ratios. However there is a way to rewrite this integral using the identity
\begin{align}
\int_{0}^{1}\,dt\, t^{n+a_1\epsilon}(1-t)^{\epsilon} = \frac{1+n-\epsilon (1+a_1+(1+n)S_{n} )} { (1+n)^2 }+O(\epsilon^2).\label{eq:eq:numeratorint2ptexpressionInt}
\end{align}
The advantage of introducing this integral is that now the sum over $n_{38}$ and $n_{78}$ can be performed effortlessly through the integral representation form of the one loop three point integral
\begin{align}
\int \frac{d^dx_7}{x_{17}^2x_{37}^2x_{47}^2} = \frac{1}{(x_{13}^2)^{3-\frac{d}{2}}}\bigg[f_0\left(\frac{x_{34}^2}{x_{13}^2},\frac{x_{14}^2}{x_{13}^2}\right)+\epsilon \, f_1\left(\frac{x_{34}^2}{x_{13}^2},\frac{x_{14}^2}{x_{13}^2}\right)\bigg]\,,
\label{3pt1loop}
\end{align}
where $f_0$ and $f_1$ are simple combination of polylogarithms recalled in equation \eqref{f0eq} of appendix \ref{app:masterintegrals}. Then each term in the series expansion in $\epsilon$ that appears in the summand can be recast in an integral representation. For example, the most complicated one, which involves the harmonic number $S_n$, can be written as 
\begin{align}
I_3 &= x_{14}^2\sum_{n,m}\frac{ S_{1+n+m} (2x_{12}\cdot x_{13})^{n}}{(1+n+m)m! }\left(-z\cdot \frac{d}{dx_1}\right)^{m} \int \frac{d^dx_7}{x_{17}^2x_{37}^2x_{47}^2 }\bigg|_{z=x_{12}}+\dots\\
=&-\int dt \frac{\ln (1-t) f_0\left(\frac{u_3 u_5}{u_4(1-t(1-u_2u_5))},\frac{t(1-u_5)}{u_4(1-t(1-u_2u_5))}\right)}{x_{13}^2 (1-t(1-u_2u_5))^2}+\dots
\end{align} 
where the $\dots$ represent the other terms in the $\epsilon$ expansion that can be handled similarly, in the end we obtain a simple integral representation for the region $I_3$, which we refrain from writing here due to its size.

The region $I_{4,2}$ can be done in a similar manner, minding some small differences. Since the one loop three point integral over $x_8$ is being evaluated exactly on the light-cone $x_{ 12}^2 = 0$, there is an extra divergence in $\epsilon$ which, in turn, requires us to expand (\ref{eq:numeratorint2ptexpression}) to the next higher order in $\epsilon$. This by no means spoils our strategy since we can still represent the extra terms of the $\epsilon$ expansion as an integral representation of the type (\ref{eq:eq:numeratorint2ptexpressionInt}).

The last region, $I_{4,4}$, computes the original integral with the constraints $x_{12}^2=x_{34}^2=0$. This effectively makes it simpler to deal with than the original integral, but this is still by far the hardest region to compute. There is simple no way to compute it using the same strategy we described above. Therefore, to proceed we used integration-by-parts identities, also known as IBP reduction \cite{Lee:2012cn,Lee:2013mka}. These identities allow us to rewrite any Feynman integral as a linear combination of a finite set of integrals, which are usually called \textit{master integrals}. The number of master integrals depends crucially on the number of loops, the number of external points, and the kinematics.  While the original integral $\cD_{24,135}$ can be expressed in terms of $74$ master integrals, the one from region $I_{4,4}$ is given by a linear combination of only $28$ master integrals
\begin{align}
	\label{eq:DintHardRegion}
	I_{4,4} = \sum_{i=1}^{28}\alpha_{i}g_i
\end{align}
where $\alpha_i$ are known rational functions of the dimension $d$ and the three cross ratios $u_2,u_4,u_5$ and $g_i$ are the master integrals, written explicitly in appendix \ref{app:masterintegrals}. The problem, then, is to evaluate these $28$ functions. 

The key point is that this system is closed: all the derivatives with respect to the remaining three cross ratios of each one of these $28$ master integrals can also be expressed as a linear combination (with rational coefficients) of the same $28$ integrals, forming then a system of first order differential equations that we need to solve, which we also attach in the auxiliary \texttt{Mathematica} file.

We were able to compute $16$ master integrals as explicit combination of multiple polylogarithms of the cross-ratios $u_2,u_4,u_5$. However, for the remaining $12$ we could not solve the differential equations while keeping arbitrary values of the cross-ratios. Instead, we considered $u_4$ and $u_5$ generic, while $u_2$ is expanded around one. This is not a random point to expand around, it is related to the OPE limit of correlators, which we treat carefully in the next section. 

The most non-trivial part of solving this system of equations is to find the solution for general $u_4$ and $u_5$ with $u_2=1$. With the dependence on two cross-ratios worked out, it is trivial to expand the differential equations as powers in the remaining cross-ratios $u_2$, and solve for each one of the master integrals $g_i$ systematically. For example, we did this up to order $\mathcal{O}((1-u_2)^{10})$ which will allow us to access plenty of perturbative data in the correlator analysis.

Finally, we can sum each one of the regions we computed and reconstruct the wanted integral $\cD_{24,135}$. In order to evaluate all the integrals that appear in the correlator, one follows the same recipe: take the integral and break it into regions. Some regions are \textit{simple} and can be computed like $I_3$, while others are \textit{hard} are expressed in terms of master integrals like $I_{4,4}$. Plugging in the computed $g_i$ computes the \textit{hard} part and summing this result with the \textit{simple} part yields the expression for the original integral. These results are in the attached \texttt{Mathematica} file.

After using such complex methods to compute these integrals it is good to perform some checks. Each one of the regions as well as the master integrals have $\epsilon$ divergences coming from dimension regularization, but once we combine then in a conformal integral all the $\epsilon$ divergences must cancel. This was essential for us to debug several steps of our computations. In the end, when possible, we also do a sanity check by comparing the results with the plane expressions from parametric integrations.


\subsection{Logarithm terms}
\label{secLog}
While the two loop integrals we are considering may be complicated, the terms that diverge with the distances that are taken to null are easier to compute. This divergent terms come only from the \textit{simple} part of the integrals, which we can explicitly integrate (after some tricks) while power law corrections come from the \textit{hard} part which are given by a combination of master integrals. If we are interested only in the leading behaviour of the correlator, which is often the case, we can neglect this \textit{hard} part and focus only in the more treatable \textit{simple} part of the integrals.

The integrands of $\mathcal{L}$, $\mathcal{S}$ and $\cD$ are rational functions of $x_{ij}^2$. Since there are two integrations over internal points and we take the limit $x_{12}^2, x_{34}^2 \to 0$, the resulting integral can have at most a quadratic divergence in logarithms. Therefore, in this regime a generic two loop conformal integral $I$ can be expressed as
\begin{equation}
	I = c_{20} \ln^2(u_1) + c_{11}\ln(u_1)\ln(u_3)
 + c_{02}\ln^2(u_3) + c_{10} \ln(u_1)+ c_{01} \ln(u_3) + c_{00}
\end{equation}
where the $c_{ij}$ are rational functions of $u_2,u_4,u_5$. In general, the expression for the integral has uniform transcendentality, which at this loop order is four. Hence, the functions multiplying logs will have lower transcendentality (for example, $c_{20}$ has transcendentality two) and thus are easier to evaluate.

In order to compute these coefficients, we turn to the method of asymptotic expansions detailed earlier. As an example, let's evaluate the coefficient $c_{20}$ of the following integral 
\begin{align}
	\mathcal{L}_{1;23;45}=&\int \frac{d^4x_7d^4x_8}{x_{17}^2x_{18}^2  x_{27}^2x_{37}^2x_{48}^2x_{78}^2x_{58}^2},,
\end{align}
in the limit $x_{12}^2 \to 0$, this integral can be expressed as the sum of the following regions
\begin{align}
	I_1=&\sum_{m_i=0}^{\infty}\frac{2^{m_1+m_2}}{\left(x_{14}^2\right)^{m_2+1}\left(x_{13}^2\right)^{m_1+1}}\int\frac{d^{d}x_{6} d^{d}x_{7} \left(x_{13}\cdot x_{16}\right)^{m_1}\left(x_{14}\cdot x_{17}\right)^{m_2}}{x_{16}^2 x_{17}^2 x_{26}^2 x_{67}^2} \\
	I_2=&\sum_{m_i=0}^{\infty}\frac{2^{m_1+m_2}}{\left(x_{13}^2\right)^{m_1+1}}\int \frac{ d^{d}x_{6} d^{d}x_{7}  \left(x_{13}\cdot x_{16}\right)^{m_1} \left(x_{16}\cdot x_{17}\right)^{m_2}}{x_{16}^2\left(x_{17}^2\right)^{m_2+2} x_{26}^2 x_{47}^2},
\end{align}
where region 3 integrates to zero and region 4 can be neglected since it does not generate $\ln{x^2_{12}}$ terms. If we wanted to compute the coefficient of other logarithm such as $c_{02}$, we could not immediately neglect these other regions. One must again perform the asymptotic expansion method of taking $x^2_{34}\to 0 $ and consider the subregions of those integrals that could generate logarithm divergences. After performing this analysis on our example, it turns out that this particular integral has no $\ln(x_{34}^2)$ dependence and therefore it has coefficients $c_{02}=c_{11}=c_{01}=0$. Thus all log dependence is encoded in the regions $I_1$ and $I_2$ written above, which can be evaluated using equation (30) of \cite{Eden:2012rr}, yielding
\begin{align}
	I_1=&\sum_{m_i=0}^{\infty}\frac{\left(2x_{13}\cdot x_{12}\right)^{m_1}\left(2x_{14}\cdot x_{12}\right)^{m_2}H(1,1,m_2,0)H(3-d/2,1,m_1+m_2,0)}{\left(x_{14}^2\right)^{m_2+1}\left(x_{13}^2\right)^{m_1+1}(x_{12}^2)^{4-d}}\\
	I_2=&\sum_{m_i=0}^{\infty}\frac{\left(2x_{13}\cdot x_{12}\right)^{m_1}\left(2x_{12}\cdot x_{14}\right)^{m_2}H(1,1,m_1+m_2,0)H(m_2+2,1,m_2,0)}{(x_{12}^2)^{2-d/2}\left(x_{13}^2\right)^{m_1+1}\left(x_{14}^2\right)^{m_2+3-d/2}}
\end{align}
where the functions $H$ are given by a ratio of gamma functions
\begin{equation}
	H(\alpha,\beta,s,i)=\frac{\Gamma (d-2) \Gamma \left(d/2+i-\beta \right) \Gamma \left(\alpha +\beta +i-d/2\right) \Gamma \left(d/2-i+s-\alpha \right)}{(2-d/2)\Gamma (\alpha ) \Gamma (\beta ) \Gamma \left(2-d/2\right) \Gamma \left(d/2-1\right)^2 \Gamma (d+s-\alpha -\beta )}.
\end{equation}

Expanding around $\epsilon=0$, it is easy to isolate the logarithm contributions, for instance the $\ln^2(x_{12}^2)$ dependence of each region is given by
\begin{align}
	I_1\sim&\sum_{m_i=0}^{\infty}\frac{\left(2x_{13}\cdot x_{12}\right)^{m_1}\left(2x_{14}\cdot x_{12}\right)^{m_2}}{(1+m_2)(1+m_1+m_2)\left(x_{14}^2\right)^{m_2+1}\left(x_{13}^2\right)^{m_1+1}}\ln^2(x_{12}^2),\\
	I_2\sim&\sum_{m_i=0}^{\infty}\frac{-\left(2x_{13}\cdot x_{12}\right)^{m_1}\left(2x_{12}\cdot x_{14}\right)^{m_2}}{2(1+m_2)(1+m_1+m_2)\left(x_{13}^2\right)^{m_1+1}\left(x_{14}^2\right)^{m_2+1}}\ln^2(x_{12}^2),
\end{align}
combining the contribution of these two integrals allows us to write the coefficient $c_{20}$ as the following sum
\begin{equation}
	c_{20} = \sum_{m_i=0}^{\infty}\frac{\left(2x_{13}\cdot x_{12}\right)^{m_1}\left(2x_{14}\cdot x_{12}\right)^{m_2}}{2(1+m_2)(1+m_1+m_2)\left(x_{14}^2\right)^{m_2+1}\left(x_{13}^2\right)^{m_1+1}}\,.
\end{equation}

By making use of the identity $(1+a)^{-1}= \int_0^1 t^a dt$ in the expression of $c_{20}$, one can recast this coefficient as a simple integral representation
\begin{equation}
	c_{20} = \int_0^1\int_0^1\frac{dt_1 dt_2}{2(x_{13}^2-2t_2x_{13}\cdot x_{12})(x_{14}^2-2t_1 t_2x_{14}\cdot x_{12})}
	\label{c20int}
\end{equation}
which can be explicitly integrated
\begin{align}
	c_{20} =& \frac{u_4}{4(1-u_5)x_{13}^2}(2\, \text{Li}_2\left(u_2\right)-2\, \text{Li}_2\left(u_2 u_5\right)+\ln \left(u_5\right){}^2+2 \ln \left(u_2\right) \ln \left(u_5\right)+\\
	-2 &\ln \left(1-u_2 u_5\right) \ln \left(u_5\right)+2\ln \left(1-u_2\right) \ln \left(u_2\right)-2 \ln \left(u_2\right) \ln \left(1-u_2 u_5\right))
	\label{c20integrated}
\end{align}
where we can see the appearance of the expected transcendentality two functions. It is then easy to take further limits of this expressions, such as more null limits or OPE limits to study the correlator, as we discuss below.

The other coefficients, such as $c_{10}$ can be calculated in a similar fashion, with the difference being that the summand will also contain harmonic numbers coming from the expansion of the gamma functions. We can, however, use \eqref{eq:eq:numeratorint2ptexpressionInt} to write the sums as integral representations and perform the integrals explicitly obtaining similar expressions as \eqref{c20integrated}. Using this method, we managed to extract the logarithm coefficients of all two loop integrals, which are also in the attached \texttt{Mathematica} file.

\part{Correlators}
\label{corr}
\begin{figure}[t]
\centering
\includegraphics[width=0.75\textwidth]{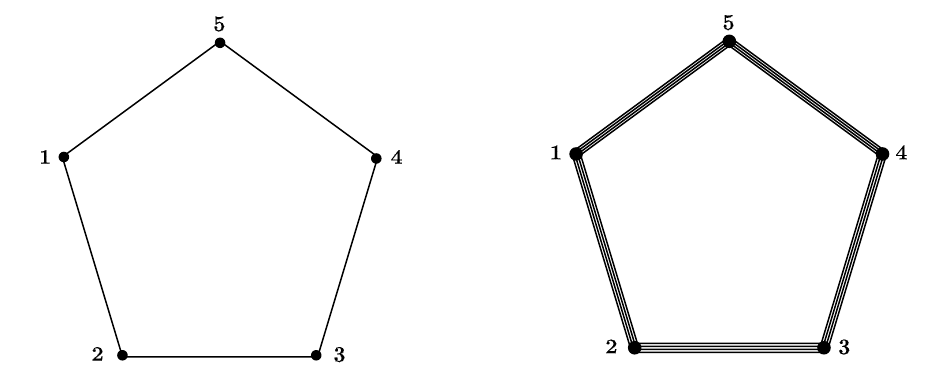}
\caption{On the left the correlation function of five weight two operators ($20^\prime$). On the right the correlation function of large weight operators, the decagon.}
\label{fig20vDec}
\end{figure}

After evaluating the relevant conformal integrals we can focus to the study of the five point correlation functions. All the cases we will study involve correlators of identical operators, which can be written as
\begin{equation}
G_5(u_1,\dots,u_5) = (x_{12}^2x_{34}^2)^{\Delta}\left(\frac{x_{15}^2x_{35}^2}{x_{13}^2}\right)^{\frac{\Delta}{2}} \langle \mathcal{O}_{\Delta}(x_1)\dots  \mathcal{O}_{\Delta}(x_5) \rangle\,.
\label{eq:5ptGeneral}
\end{equation}
where $\Delta$ is their scale dimension and $u_i$ are cross-ratios defined in \eqref{eq:conformalCrossratios}. We will focus on two particular choices of charge for the external operators in $\mathcal{N}=4$ SYM: first when the external operators are the lightest ($\Delta = 2$) and second when the external operators are heavy ($\Delta \to \infty$), as depicted in figure \ref{fig20vDec}. Physically, these two correlators are very different. The later forms a thick perimeter of propagators that effectively decouples space into two regions: inside and outside \cite{Coronado:2018cxj}. Thus, the correlator is given by the square of a more fundamental object, the \textit{decagon}, see \cite{Fleury:2020ykw,Bargheer:2018jvq}. Meanwhile the light correlator sits in the opposite spectrum of correlators and in perturbation theory maximally couples these inside and outside regions. 

The next two sections deal with these very different objects. In section \ref{secLight} we study the lightest correlator under several kinematical limits that are governed by small and large spin dynamics. In both cases, we compare our results with the literature, the small spins being compared with the perturbative computations \cite{Bianchi:2019jpy} and the large spin being matched with the constraints coming from analytic conformal bootstrap \cite{Bercini:2020msp}. In section \ref{secHeavy} we analyse the decagon, by checking its physical properties such as Steinmann relations, null limit constrains like stampedes \cite{Olivucci:2022aza} and the plane kinematics results \cite{Fleury:2020ykw}.

\section{Lightest correlator}
\label{secLight}
Any five point correlation function in a conformal field theory can be decomposed into a product of kinematical and dynamical terms
\begin{equation}
G_5(u_1,\dots,u_5) = \sum_{J_1,J_2,\ell} P_{J_1J_2\ell}
\,\mathcal{F}(u_1,\dots,u_5) \,,
\label{eq:5ptexpansion}
\end{equation}
where $\mathcal{F}(u_i)$ is the kinematical function known as the conformal block which is fully fixed by conformal symmetry, and $P_{J_1J_2\ell}$ is the OPE data that encodes all the theory dependent information of the theory.

\begin{figure}[t]
\centering
\includegraphics[width=\textwidth]{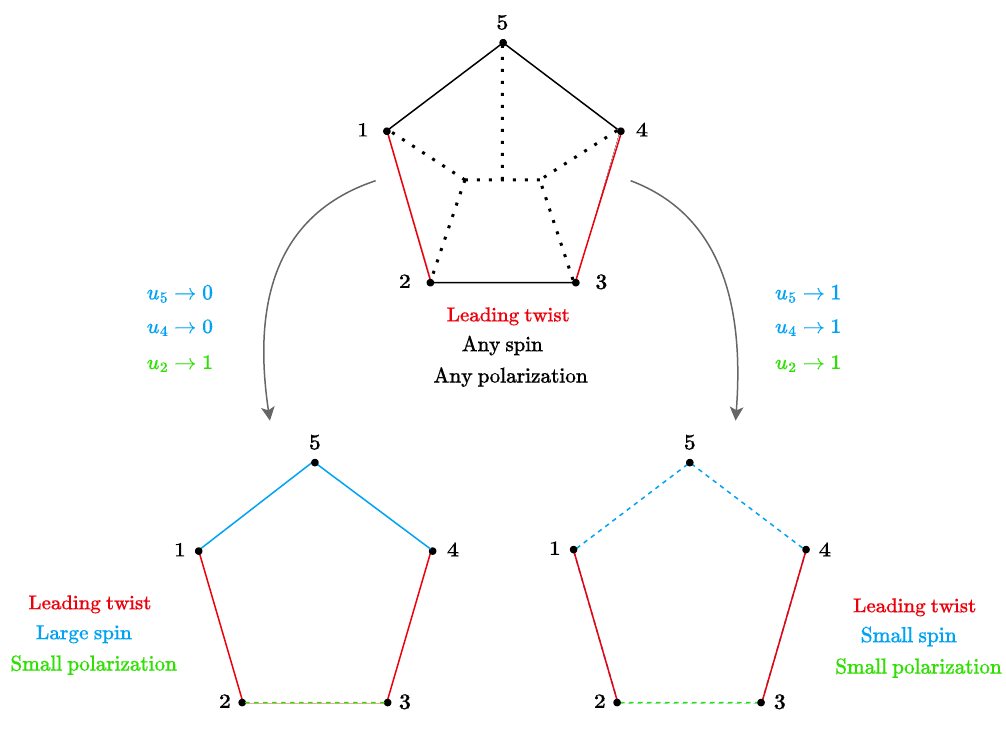}
\caption{Two kinematical limits of the correlator. Black lines represent general kinematics while solid coloured lines represent Lorenzian OPE (points become null separated) and dashed lines represent Euclidean OPE (points approach each other). The dotted lines on top represent the OPE channels we chose to decompose the correlator while the colors indicate the physical implication of the limits for the operators being exchanged. We arrive in both limits by first considering the single light-cone OPE displayed on the top.}
\label{figChannels}
\end{figure}

Even though it is completely fixed, the five point conformal block for general kinematics is a complicated object not known in closed form, \cite{Goncalves:2019znr,Poland:2023vpn,Poland:2023bny}. In order to study these higher point functions one usually turns to kinematical limits of the blocks where several simplifications occur. Here will be no different. We will start by considering the light-cone OPE between points 12 and points 34, see top of figure \ref{figChannels} (in terms of cross-ratios $u_1,u_3 \to 0$). As explained in \cite{Bercini:2020msp}, this limit projects into leading twist operators being exchanged in the OPE channels and allows to write the conformal block as an integral representation 
\begin{align}
\mathcal{F}(u_i) &= \frac{\Gamma\left(2J_1+\tau_1\right)\Gamma\left(2J_2+\tau_2\right)}{\Gamma\left(J_1+\frac{\tau_1}{2}\right)^2\Gamma\left(J_2+\frac{\tau_2}{2}\right)^2}u_1^{\frac{\tau_1}{2}}u_3^{\frac{\tau_2}{2}}u_4^{\frac{\tau_1-\Delta}{2}+\ell}u_5^{\frac{\tau_2}{2}+\ell}(1-u_2)^\ell\times
\label{LCConfBlock} \\
&\times \int_0^1\int_0^1 dt_1 dt_2 (t_1(1-t_1))^{J_1-1+\frac{\tau_1}{2}}(t_2(1-t_2))^{J_2-1+\frac{\tau_2}{2}}\times \nonumber \\
& \times \frac{(1-u_4-t_1(1-u_4-u_5+u_2u_4u_5))^{J_2-\ell}(1-u_5-t_2(1-u_4-u_5+u_2u_4u_5))^{J_1-\ell}}{(1-t_1(1-u_5)-t_2(1-u_4)+t_1t_2(1-u_4-u_5+u_2u_4u_5))^{J_1+J_2-\frac{\Delta-\tau_1-\tau_2}{2}}}\nonumber
\end{align}
where $\{J_1,\tau_1\}$ and $\{J_2,\tau_2\}$ parametrize the spin and the twist of the operators being exchanged in the channels 12 and 34, respectively.

With the conformal block defined we can turn to the dynamical object appearing in the OPE decomposition, $P_{J_1J_2\ell}$. As can be seen in figure \ref{figChannels}, this quantity is given by a combination of structure constants of one and two spinning operators 
\begin{equation}
P_{J_1J_2\ell} = C^{\bullet\circ\circ}_{J_1}C^{\bullet\circ\circ}_{J_2}C^{\bullet\bullet\circ}_{J_1,J_2,\ell}
\label{eqPtoC}
\end{equation}
where the polarization $\ell$ accounts for the several tensor structures \cite{Costa:2011mg} that appear in the three point function of two spinning operators recalled in appendix \ref{appData}.

From now on we will consider the case where the external operators are the lightest operators ($\Delta=2$), the ubiquitous $20^\prime$ operators of $\mathcal{N}=4$ SYM. The five point function integrand of such operators at two loops was already worked out in \cite{Bargheer:2022sfd} and now we can study its integrated properties. The plan is very simple, compare the integrated correlator with the block expansion in several kinematic limits and extract physical information about the structure constants of two spinning operators.


\subsection*{Finite spin}
\label{secFinSpin}
After taking the single light-cone limit we consider the Euclidean OPE between the remaining points of the five point correlator, as depicted in the right of figure \ref{figChannels}. This  projects into small spin and small polarizations being exchanged in the operator product expansion. For the cross-ratios, this limit amounts to start with $u_1,u_3\to 0$ and then take the remaining $u_i \to 1$. In the conformal blocks side, one can easily expand the integrand of \eqref{LCConfBlock} and trivially perform the integration in the auxiliary variables $t_i$. On the other hand, with the conformal integrals now evaluated we can also compute the integrated correlator at two loops. By matching these two expressions we can just read off the structure constants \eqref{eqPtoC}.

Although perfect valid, the method described above is inefficient to extract huge amounts of data. What in turn is used is the Casimir extraction method, described in Appendix E of \cite{Bercini:2021jti}. In a nutshell, this method consist of iteratively apply the quadratic Casimir on the correlator and use the fact that the conformal blocks are eigenfunctions of this differential operator to systematically read off the structure constants of the correlator.

We were able to extract more than a thousand new structure constants between two spinning operators. For polarizations $\ell=0,\dots,6$ we extracted data\footnote{For $\ell=0$ the integrals are simpler and we can extract a lot of data, we went up to $J=50$ but it is easier to generate much more. To illustrate this, in the file we also attach data for $J_1=2$ and $J_2=0,\dots,200$.} for two spinning operators up to spin $J=36$. At this point the two loop data is give by the rational numbers built from integers with almost 100 digits. This data is attached in the supplemental \texttt{Mathematica} file, and examples are
\begin{align*}
\hat{C}^{\bullet\bullet\circ}_{8,12,3} &= 1 - \lambda \left(\frac{48563133305437}{2650343295600}\right)+\lambda^2\left(\frac{359608758568933328334171885613}{1376766638568246319954560000} + \frac{1526}{165} \zeta_3\right) \\
\hat{C}^{\bullet\bullet\circ}_{10,10,4} &= 1 - \lambda \left(\frac{793855867937}{49997493000}\right)+\lambda^2\left(\frac{286981761743641772075609}{1333199630018692800000}\right)\,,
\end{align*}
where $\hat{C}$ is the tree-level normalized structure constant recalled in appendix \ref{appData}.

Despite the abundant data, we were not able to guess a closed expression for the structure constant of two spinning operators at two loops. However, we observed that the coefficient of $\zeta_3$ in the data is independent of the polarization and follows a simple pattern
\begin{equation}
\hat{C}^{\bullet\bullet\circ}_{J_1,J_2,\ell} = \dots + 24\zeta_3\lambda^2|S_1(J_1)-S_1(J_2)| +\dots 
\label{CZeta3Part}
\end{equation}
which was previously observed in an independent computation by Bianchi in \cite{Bianchi:2019jpy} and suggests the following naive large spin limit
\begin{equation}
\hat{C}^{\bullet\bullet\circ}_{J_1,J_2,\ell} \xrightarrow{\text{naive}} \dots + 24\zeta_3\lambda^2|\ln(J_1)-\ln(J_2)|+\dots
\label{CZeta3PartLarge}
\end{equation}
which is drastically different than the large spin and polarization expression bootstrapped in \cite{Bercini:2020msp}
\begin{equation}
\hat{C}^{\bullet\bullet\circ}_{J_1,J_2,\ell} \xrightarrow{\text{correct}}  \mathcal{N}e^{-\frac{f(\lambda)}{4}(\ln^2\ell +\ln 4 \ln(J_1,J_2))-\frac{g(\lambda)}{2}\ln\ell}\,.
\label{eqLargeJL}
\end{equation}

The correct expression is analytic in spins and polarizations, while the naive expression depends non-analytically as the modulus of the spins. Below we clean this apparent paradox by tracking how the structure constants go from small to large values of spin and understanding how such non-analytic contributions get washed way in the large spin limit.

\subsection*{Large spin}
\label{secLargeSpin}
Here we consider the limit depicted on the left of figure \ref{figChannels}, which we will denote by \textit{open box} limit. After taking the sigle light-cone limit we then send the distances between the points 15 and 45 to be null, thus projecting into large spin operators being exchanged. Lastly, we make points 23 approach each other projecting further into small polarizations, see \cite{Bercini:2020msp}. In terms of cross-ratios the open box limits consists in taking $u_1,u_3 \to 0$, then $u_4,u_5\to 0$ and finally $u_2 \to 1$.

More precisely, in this limit the spins are going to infinity and the cross-ratios are vanishing, but the quantities $x =2J_2\sqrt{u_4}$ and $y=2J_1\sqrt{u_5}$ are held fixed. The block in this limit was computed in \cite{Bercini:2020msp}, and it reads:
\begin{align}
\mathcal{F}(u_1,\dots,u_5) &= 2^{2(2+J_1+J_2)+\gamma_1+\gamma+2}u_1^{\tau_1/2}u_3^{\tau_2/2}(1-u_2)^\ell u_4^{\frac{2\ell+\gamma_1}{4}}u_5^{\frac{4+2\ell+\gamma_2}{4}}\times\nonumber \\
& \times K_{\ell+\frac{\gamma_1}{2}}\left(2J_2\sqrt{u_4}\right) K_{\ell+\frac{\gamma_2}{2}}\left(2J_1\sqrt{u_5}\right)\,.
\label{boxBlock}
\end{align}

It is then trivial to write the five point correlation function \eqref{eq:5ptexpansion} as
\begin{align}
G_5(u_1,\dots,u_5) &= \left(\frac{u_1 u_3 }{u_4}\right)\sum_{\ell=0}^{\infty}\int_{0}^{\infty} dx dy \left(\frac{u_1^{\frac{\gamma_1}{2}}u_3^{\frac{\gamma_2}{2}}u_4^{\frac{\gamma_1}{4}}u_5^{\frac{\gamma_2}{4}}}{2^{2\ell-\gamma_1-\gamma_2}}\frac{(1-u_2)^\ell}{\Gamma(1+\ell)^2}\right)\times\nonumber\\
&\times \hat{P}\left(\frac{x}{2\sqrt{u_4}},\frac{y}{2\sqrt{u_5}},\ell\right) x^{1+\ell}K_{\ell+\frac{\gamma_1}{2}}(x) y^{1+\ell}K_{\ell+\frac{\gamma_2}{2}}(y) 
\end{align}
where we traded sum over spins by integrals and factored out the tree-level value of the large spin structure constants
\begin{equation}
P_{J_1J_2\ell} \to \frac{2^{-2J_1-2J_2}\pi J_1^{\ell+\frac{1}{2}}J_2^{\ell+\frac{1}{2}}}{\Gamma(\ell+1)^2}\hat{P}(J_1,J_2,\ell)
\end{equation}
where the quantity $\hat{P}$ is the tree-level normalized combination of structure constants that encodes all the perturbative corrections to the OPE data, and $\gamma_i$ is the anomalous dimension of the twist two operators at large spin \cite{Drummond:2007aua},
\begin{equation}
\gamma(J) = f(\lambda)\ln{J} + g(\lambda)
\end{equation}
where $f(\lambda)$ and $g(\lambda)$ are the cusp and collinear anomalous dimensions that govern the large spin dynamics.

Following the same ideas of \cite{Alday:2013cwa,Bercini:2020msp} we assume that in the large spin limit the structure constants have the form
\begin{equation}
\hat{P}(J_1,J_2,\ell) = \sum_{a,b,c}p_\ell(a,b,c)\lambda^a\ln^b{J_1}\ln^c{J_2}
\end{equation}
where $p_\ell(a,b,c)$ are unknown coefficients. Using the conformal integrals computed here, we can evaluate the correlator at two loops in the open box limit and compare with the block expansion. Doing that fixes these coefficients and yields the following expression for the large spin, small polarization structure constants of two spinning operators
\begin{equation}
C^{\bullet\bullet\circ}_{J_1,J_2,\ell} = \mathcal{N}_\ell e^{-\frac{\ln{2}}{2}f(\lambda)\ln{J_1J_2}}
\label{largeC}
\end{equation}
of course the expression above can only be trusted at two loops, but it is tempting to conjecture the all loop expression above since our expressions have the same spin dependence as the large spin and large polarizations results derived in \cite{Bercini:2020msp} and presented in \eqref{eqLargeJL}. The only difference is that the large polarization information gets washed away into the normalization $\mathcal{N}_\ell$. The bootstrap arguments of \cite{Bercini:2020msp} could not fix the spin-independent normalization. However, our direct comparison allows us to explicitly compute it to be
\begin{equation}
\mathcal{N}_{\ell} =\left(2^{-g(\lambda)}e^{-2\gamma_E g(\lambda)+\gamma_E^2 f(\lambda)-8\zeta_2\lambda+96\zeta_4\lambda^2}\right)\times \left( e^{-2(S_1(\ell)^2-S_2(\ell))\lambda+\lambda^2 n_\ell}\right)
\label{normalization}
\end{equation}

The normalization is given by the product of two terms. The first is polarization independent while the second depends on the polarization $\ell$. At one loop we were able to express this dependence as some simple harmonic numbers, but at two loops we could not find an analogous expression\footnote{Where $\gamma_E$ is the Euler's constant and the $n_\ell$ are a rational functions that we can explicit compute but could not find a pattern for: $n_0 = 0$, $n_1 = 22$, $n_2=303/8$, $n_3=32639/648$ $\dots$.}. Opposed to the structure constants, we could also not write the normalization factor in a way that hints a lift to an all loop expression. However, at one loop where we do have an analytic expression in terms of polarizations we can do a nice sanity check: by further taking the limit $\ell \to \infty $ in \eqref{largeC} we can recover the bootstraped result of \cite{Bercini:2020msp} for large spins and polarizations.

At this point we turn to the apparent contradiction between the correct (\ref{largeC}) and the naive \eqref{CZeta3Part} expressions for the large spin finite polarization structure constants
\begin{align}
\hat{C}^{\bullet\bullet\circ}_{J_1,J_2,\ell} &\xrightarrow{\text{naive}} \dots + 24\zeta_3\lambda^2|\ln(J_1)-\ln(J_2)|+\dots \\
\hat{C}^{\bullet\bullet\circ}_{J_1,J_2,\ell} &\xrightarrow{\text{correct}} \dots + 24\zeta_3\lambda^2(2\gamma_{E}-\ln{2})+\dots
\end{align}

Since this absolute value feature is shared among all finite polarizations, let's focus in the simplest case of $\ell=0$, where it is sufficient to consider the correlator at $u_2=1$. In this regime, the two loop correlator can be written as the sum of two terms
\begin{equation}
G_5 =1 +\lambda(\dots)+ \lambda^2\left(\zeta_3\frac{u_5(1+u_5)(1-u_4)^3\ln{u_4}-u_4(1+u_4)(1-u_5)^3\ln{u_5}}{(u_4-u_5)(1-u_4u_5)}+R\right)
\label{zetaBreak}
\end{equation}
where the first term has an explicit $\zeta_3$ dependence and the second term is a complicated combination of multiple polylogarithms, which in principle have no $\zeta_3$ dependence
\begin{equation}
R =-3(9+u_4)(1+u_5)\text{Li}_{1}(1-u_4)\text{Li}_{3}(1-u_4)+12(1-u_4)(1+u_5)\text{Li}_{1}(1-u_5)\text{Li}_{3}(1-u_4)+\dots
\label{Rgeneral}
\end{equation}
and we ignore the one loop part for our analysis since it has no $\zeta_3$ and it has no problems whatsoever when analytic continuing to large spins.

When we OPE expand the expression above, namely when we expand $u_4$ and $u_5$ around one, the polylogarithms in the $R$-term indeed never generate a factor with $\zeta_3$. All the $\zeta_3$ dependence of the structure constant comes from the first term of (\ref{zetaBreak}). Moreover, by comparing with the block expansion one can explicitly check that the OPE expansion of this term indeed generates the observed $\zeta_3 |S_1(J_1)-S_1(J_2)|$ behaviour of the data.

Now let's consider what happens at large spin, or in terms of cross-ratios when we take $u_4\to0$ and $u_5\to0$. First, due to its denominator, the first term in \eqref{zetaBreak} has an order of limit issue, which can be understood as the large spin incarnation of the non-analytical absolute value. However, in the limit of the cross-ratios going to zero the multiple polylogarithms in the $R$ part do give a $\zeta_3$ dependence, for example the two terms written in \eqref{Rgeneral} give
\begin{equation}
R =27\zeta_3\ln{u_4}-12\zeta_3\ln{u_5}+\dots
\label{Rlarge}
\end{equation}
combining all the new $\zeta_3$ dependence that arises in the null limit of the $R$ term with the starting $\zeta_3$ dependence of the two loop correlator precisely cancels the order of limits issue and yields a simple open-box limit of the correlator
\begin{equation}
G_5 = -12\zeta_3(\ln{u_1}+\ln{u_3}+\ln{u_4}+\ln{u_5})
\end{equation}
which in turn gets translated to the simple large spin behaviour of the structure constants observed in (\ref{largeC}).

This simple two loop analysis  of spinning OPE coefficients proved to be crucial to clarify a potential problem in the assumptions that are usually done in analytical conformal bootstrap studies \cite{Alday:2013cwa,Bercini:2020msp,Bercini:2021jti}. Obviously, it would be even better to put these assumptions on a firmer footing by deriving a generalization of the, by now well established, Lorentzian inversion formula \cite{Caron-Huot:2017vep}. On the other hand it also gives motivation to delve into the extremely non-trivial case of two loop six point functions which we leave to the future.  

Another interesting kinematical limit of the correlator is the null polygon limit, where we take all the cross-ratios to zero. Following the procedure of \ref{secLog} we were able to evaluate the log divergences of the conformal integrals, and consequently compute the leading behaviour of the correlator, it reads
\begin{equation}
\ln{G_5} = \mathcal{C}_0 + \sum_{i=1}^{5} (-2\lambda+4\lambda^2\zeta_2)\ln{u_i}\ln{u_{i+1}}+4\zeta_2\lambda^2(\ln{u_i}\ln{u_{i+2}}+\ln^2{u_i})-12\zeta_3\lambda^2\ln{u_i}
\label{nullC}
\end{equation}
being the constant $\mathcal{C}_0 = -8\zeta_2\lambda+126\zeta_4\lambda^2$. This expression for the null correlator is the same as the two loop expansion of the bootstrapped result presented in (17) of \cite{Bercini:2020msp}, up to the constant $\mathcal{C}_0$, which could not be fixed by bootstrap arguments.

In the null polygon limit all the cross-ratios are going to zero, while both spins and polarizations are going to infinity. Under these circumstances the conformal block exponentiates \cite{Bercini:2020msp} and we can write the five point correlator in a very simple form 
\begin{align}
G_5(u_i) &= \frac{u_1 u_3}{u_2 u_4 u_5} \int dJ_1 dJ_2 d\ell\, \frac{u_1^{\frac{\gamma_1}{2}}u_3^{\frac{\gamma_2}{2}}}{2^{-2-\gamma_1-\gamma_2}} \frac{J_1^{1+\frac{\gamma_2}{2}}J_2^{1+\frac{\gamma_1}{2}}}{\ell^{2-\frac{\gamma_1}{2}-\frac{\gamma_2}{2}}}\hat{P}(J_1,J_2,\ell)\,e^{-\ell u_2 -\frac{J_2^2}{\ell}u_4-\frac{J_1^2}{\ell}u_5}\,,
\label{blockPent}
\end{align}
in order to study the structure constants for large spin and large polarizations we again start by making an ansatz for the expression of these OPE coefficients
\begin{equation}
\hat{P}(J_1,J_2,\ell) = \sum_{a,b,c}p(a,b,c,d)\lambda^a\ln^b{J_1}\ln^c{J_2}\ln^d{\ell}\,,
\end{equation}
and by comparing the block expansion \eqref{blockPent} with the null correlator \eqref{nullC} we can fix all these $p$-coefficients. In the end we recover the bootstrapped result of \cite{Bercini:2020msp} for the large spin and polarization structure constants presented in \eqref{largeC} being the only difference that now we have access to the overall normalization of the structure constants
\begin{equation}
\mathcal{N}=2^{-g(\lambda)}e^{-8\zeta_2\lambda+101\zeta_4\lambda^2+\frac{5\gamma_E}{4}(\gamma_E f(\lambda)-2g(\lambda))}
\end{equation}
which is correct up to two loops, but we also were not able to write in a way that hints a lift to an all loop expression.

\section{Large weight correlator}
\label{secHeavy}
\begin{figure}
 	\centering
 	\includegraphics[width=0.75\linewidth]{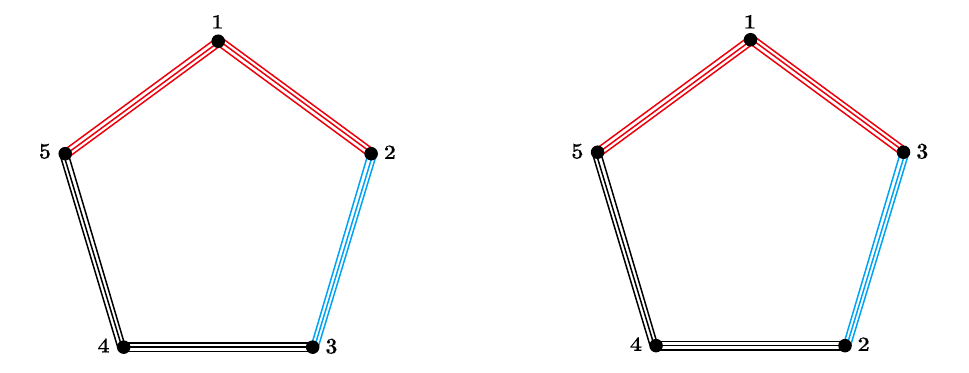}
 	\caption{ On the left the Decagon with canonical ordering and on the right the Decagon with non-canonical ordering of the operators. The red lines represent $X$ propagators, the blue lines $Y$ propagators and the black lines $Z$ propagators.}
 	\label{fig:decagonpicture}
 \end{figure}

Now we shift gears and consider correlators built by large charge operators. We start by picking specific R-charge polarizations for the operators in the five point correlator and define the decagon as the large charge limit ($k\to \infty$) of the following correlation function
\begin{align}
\langle \textrm{tr} (X^{2k})(x_1)\textrm{tr}(\bar{X}^{k}\bar{Y}^{k} )(x_2) \textrm{tr}(\bar{Z}^{k}Y^{k} )(x_3)\textrm{tr}(Z^{2k} )(x_4)\textrm{tr}(\bar{Z}^{k} \bar{X}^{k} )(x_5)  \rangle  = \frac{\mathbb{D}^2(u_i)}{(x_{12}^2x_{23}^2x_{34}^2x_{45}^2x_{15}^2)^k}
\end{align} 
where $X,Y, Z$ and their bar counterparts are the six scalar operators of $\mathcal{N}=4$ SYM, see figure \ref{fig:decagonpicture}. This is the five point generalization of the famous simplest four point correlation function
\begin{align}
\langle \textrm{tr}(\bar{Z}^{k} \bar{X}^{k} )(x_1) \textrm{tr}( X^{2k} )(x_2)  \textrm{tr}(\bar{Z}^{k} \bar{X}^{k} )(x_3)\textrm{tr}(Z^{2k}  )(x_4) \rangle = \frac{\mathbb{O}^2(z,\bar{z})}{(x_{12}^2x_{23}^2x_{34}^2x_{14}^2)}\,,
\end{align}
introduced in \cite{Coronado:2018cxj} and bootstraped by demanding the following three axioms
\begin{itemize}
\item The basis of functions are built from four point ladder integrals.
\item There is log truncation in the double trace OPE channel (Steinmann relations).
\item Constraints from the null square limit.
\end{itemize} 

The decagon, on the other hand, cannot be bootstrapped at arbitrary loop orders since the very first step used in the octagon fails: at high loop orders the basis of functions that govern the decagon is unknown, making it impossible to systematically study the constrains coming from OPE and null limits. However, at two loops we have all the ingredients to study the decagon. A simple ansatz inspired by the ideas of \cite{Bargheer:2022sfd} together with result of the plane \cite{Fleury:2020ykw} is enough to obtain the integrand of the decagon\footnote{By cyclic permutations we mean taking $u_i \to u_{i+1}$ and shift the indices of the integrals by one unity in a cyclic way, as presented in \eqref{intLStein} and \eqref{intSStein}.}  for general kinematics
\begin{align}
&\mathbb{D}=1-\lambda \left(1+u_1 u_4-u_5\right) \Phi^{(1)}\left(u_1 u_4,u_5\right)+2\lambda^2\left(2\left(1+u_1 u_4-u_5\right) \Phi^{(2)}\left(u_1 u_4,u_5\right)+\right.\nonumber\\
&\left.+\left(u_1u_4+u_2u_5-u_1-u_5\right) \mathbb{L}_{1,23,45}+\left(1-u_3+u_2u_4\right) \mathbb{S}_{1,25,34}\right)+\textrm{cyclic}+O(\lambda^6)
\label{decResult}
\end{align}
being and $\Phi^{(L)}$ the well-known ladder integrals \cite{Usyukina:1993ch} reminded in appendix \ref{appInts} and the other conformal integrals are recalled below
\begin{align}
    \mathbb{L}_{\textcolor{c1}{1};\textcolor{c2}{2}\textcolor{c3}{3};\textcolor{c4}{4}\textcolor{c5}{5}} &= x_{\textcolor{c1}{1}\textcolor{c3}{3}}^2x_{\textcolor{c1}{1}\textcolor{c4}{4}}^2x_{\textcolor{c2}{2}\textcolor{c5}{5}}^2\int \frac{d^4x_7d^4x_8}{x_{\textcolor{c1}{1}7}^2x_{\textcolor{c1}{1}8}^2  x_{\textcolor{c2}{2}7}^2x_{\textcolor{c3}{3}7}^2x_{\textcolor{c4}{4}8}^2x_{\textcolor{c5}{5}8}^2x_{78}^2}\\
    \mathbb{S}_{\textcolor{c1}{1},\textcolor{c2}{2}\textcolor{c5}{5},\textcolor{c3}{3}\textcolor{c4}{4}} &= x_{\textcolor{c2}{2}\textcolor{c4}{4}}^2x_{\textcolor{c2}{2}\textcolor{c5}{5}}^2x_{\textcolor{c3}{3}\textcolor{c5}{5}}^2\int \frac{d^4x_7\, d^4x_8 x_{\textcolor{c1}{1}7}^2}{x_{\textcolor{c1}{1}8}^2 x_{\textcolor{c2}{2}7}^2x_{\textcolor{c5}{5}7}^2x_{\textcolor{c5}{5}8}^2 x_{\textcolor{c3}{3}7}^2 x_{\textcolor{c4}{4}8}^2 x_{\textcolor{c4}{4}7}^2x_{78}^2}\,.
\end{align}
    
We will take advantage of the integrals that were computed in the first part of the paper to verify that the decagon \eqref{decResult} satisfies the last two bootstrap axioms listed above. After confirming that the axioms hold, we can also invert the logic and ask: how much of the decagon one can fix by demanding the bootstrap constrains? This is what we turn to next.

\subsection*{Steinmann}

The second axiom states that the octagon, when expanded in the double trace OPE, is linear in logs of the cross ratios that are approaching zero. This property can be equivalently recast in terms of Steinmann relations that appear frequently in scattering amplitudes. This property is easier to express in terms of dual momenta variables $p_i\equiv x_{i}-x_{i+1}$, meanwhile the usual Mandelstam invariants $s_{ij}=(p_i+p_j)^2$ can be written in terms of $x_{i}$ as
\begin{align}
s_{12}=x_{13}^2 ,\ \ \  s_{23}=x_{24}^2,\ \ \  s_{34}=x_{35}^2,\ \ \  s_{45}=x_{14}^2 ,\ \ \  s_{15}=x_{25}^2
\end{align}

Steinmann relation is the statement that \textit{the double discontinuity in overlapping channels is zero}. For example, the action of 
 $\textrm{disc}_{s_{12}}\textrm{disc}_{s_{23}}$ on the decagon should vanish. 
Equivalently, it states that there should be no $\ln x_{13}^2 \ln x_{24}^2$ in the decagon\footnote{Note that the limit $x_{13}^2,x_{24}^2\rightarrow 0$ can be mapped, under permutation, to the limit $x_{12}^2,x_{34}^2\rightarrow 0$ discussed in the previous sections or equivalently we could have used non-conventional labelling of the poitns depicted on the right of figure \ref{fig:decagonpicture} where Steinmann would translate into the absence of $\ln u_1 \ln u_3$ in the decagon in the limit $u_1,u_3\rightarrow 0$.} in the limit $x_{13}^2,x_{24}^2\rightarrow 0$. 

Besides the four point ladder integrals which are already known to satisfy Steinmann relations, the only five point integrals that satisfy the Steinmann relation independently  are
\begin{align}
&\mathbb{L}_{\textcolor{c1}{1},\textcolor{c2}{2}\textcolor{c3}{3},\textcolor{c4}{4}\textcolor{c5}{5}},  \ \ \ \mathbb{L}_{\textcolor{c2}{2},\textcolor{c1}{1}\textcolor{c5}{5},\textcolor{c3}{3}\textcolor{c4}{4}},   \ \ \ \mathbb{L}_{\textcolor{c3}{3},\textcolor{c1}{1}\textcolor{c2}{2},\textcolor{c4}{4}\textcolor{c5}{5}},  \ \ \  \mathbb{L}_{\textcolor{c4}{4},\textcolor{c1}{1}\textcolor{c5}{5},\textcolor{c2}{2}\textcolor{c3}{3}},  \ \ \ \mathbb{L}_{\textcolor{c5}{5},\textcolor{c1}{1}\textcolor{c2}{2},\textcolor{c3}{3}\textcolor{c4}{4}}\label{intLStein}\\
&\mathbb{S}_{\textcolor{c1}{1},\textcolor{c2}{2}\textcolor{c5}{5},\textcolor{c3}{3}\textcolor{c4}{4}}, \ \ \  \mathbb{S}_{\textcolor{c2}{2},\textcolor{c1}{1}\textcolor{c3}{3},\textcolor{c4}{4}\textcolor{c5}{5}}, \ \ \ \ \mathbb{S}_{\textcolor{c3}{3},\textcolor{c2}{2}\textcolor{c4}{4},\textcolor{c1}{1}\textcolor{c5}{5}},  \ \ \ \mathbb{S}_{\textcolor{c4}{4},\textcolor{c3}{3}\textcolor{c5}{5},\textcolor{c1}{1}\textcolor{c2}{2}}, \ \ \ \  \mathbb{S}_{\textcolor{c5}{5},\textcolor{c1}{1}\textcolor{c4}{4},\textcolor{c2}{2}\textcolor{c3}{3}}\label{intSStein} \,,
\end{align}
which are precisely the only integrals that appear in the plane computation of \cite{Fleury:2020ykw}. One could imagine that some strange combination of integrals with rational prefactors could, in principle, satisfy Steinmann relations. At leading order in the OPE expansion this is indeed the case, but by requiring absence of double discontinuities at every order in the OPE expansion makes it so that no such combination exists\footnote{We can explicitly check this in the case of the two loop five point integrals here considered}.  Therefore, only the self-satisfying Steinmann integrals of \eqref{intLStein} and \eqref{intSStein} can contribute.

This two loop fulfillment of Steinmann relations also suggests that they might be satisfied at all loops for the decagon, just as is the case for the octagon. It would be interesting to use this property to severely reduce the possible integrals that can contribute to the decagon at higher loops. Although difficult, this task could be treatable, since these log terms of the integrals are precisely the easiest ones to evaluate using the asymptotic expansion method of section \ref{secLog}.

\subsection*{Stampedes}
The third axiom for the octagon can be recast in the so-called stampedes constraints \cite{Olivucci:2021pss}, which were later extended for other polygons such as the decagon \cite{Olivucci:2022aza}. The stampedes limit is the combination of the null pentagon limit $u_i\rightarrow 0$ with perturbation theory, $\lambda \to 0$, where the \textit{cusp times} are held fixed
\begin{equation}
t_k^2=-\lambda\log{u_{k-1}}\log{u_{k}}\,.
\end{equation}
When these special polygons (octagon, decagon, etc..) do not have internal diagonals, the stampedes limit predicts an exponentiation. For instance, the decagon is conjectured to be
\begin{align}
\mathbb{D} = e^{-(t_1^2+t_2^2+t_3^2+t_4^2+t_5^2)}\,.
\label{StampedesPrediction}
\end{align}

The leading log behavior can be easily computed via the asymptotic expansions methods of section \ref{secLog}, and gives the following expression for the stampedes limits of the integrals
\begin{align}
\lambda\Phi^{(1)}(u_1u_4,u_5) &= -(t_1^2 +t_5^2) \\
\lambda^2\Phi^{(2)}(u_1u_4,u_5) &= \frac{(t_1^2+t_5^2)^2}{4}\\
\lambda^2\mathbb{L}_{1,23,45} &=\frac{t_1^4}{4}+\frac{t_1^2t_2^2}{2}+\frac{t_1^2t_5^2}{2}+t_2^2t_5^2 \\ 
\lambda^2\mathbb{S}_{1,25,34} &=\frac{t_2^2t_3^2}{2}+ \frac{t_4^2t_5^2}{2}+t_1^2t_4^2+t_1^2t_3^2
\end{align}
It is then trivial to plug these results back into the expression for the decagon \eqref{decResult} and check that they give the same as the two loop expansion of the stampedes result \eqref{StampedesPrediction}. 

Similar to Steinmann relations, the stampedes also depend only on easier to compute log terms. Both constraints are expected to hold at all loops, which motivates the bootstrap analysis of how constraining are these conditions.

\subsection*{Bootstrap}
Any five point correlator \eqref{eq:Gnloop} can be written as a combination of the two loop four and five point integrals: $\Phi^{(L)}$, $\mathcal{L}_{i,jk,lm}$, $\mathcal{S}_{i,jk,lm}$ and $\mathcal{D}_{ij,klm}$, see \cite{Bargheer:2022sfd,ToAppear}. By demanding Steinmann relations we severely reduce the basis of integrals to be only four point integrals, \eqref{intLStein} and \eqref{intSStein}. At this point one can write that the most general two loop decagon function is given by
\begin{equation}
    \mathbb{D}_{\texttt{2-loops}}=\lambda^2\left(a_1(u_i)\Phi^{(2)}\left(u_1 u_4,u_5\right)+b_1(u_i)\mathbb{L}_{1,23,45}+c_1(u_i) \mathbb{S}_{1,25,34}\right)+\textrm{cyclic}
    \label{decAnsatz}
\end{equation}
where $a_i,b_i,c_i$ are unknown rational functions of the cross-ratios related to each other by cyclic symmetry
\begin{equation}
     a_i|_{u_i \to u_{i+1}} = a_{i+1}\,,\quad b_i|_{u_i \to u_{i+1}} = b_{i+1}\,,\quad c_i|_{u_i \to u_{i+1}} = c_{i+1}\,.
\end{equation}

Demanding that the putative decagon \eqref{decAnsatz} matches the stampedes prediction \eqref{StampedesPrediction} completely fixes the null limit of these polynomials
\begin{equation}
a_i(u_k\to 0) = 4, \quad b_i(u_k\to 0) = 0, \quad c_i(u_k\to 0) = 2\,.
\label{stampPoly}
\end{equation}
where the vanishing of the $b_i$ coefficients was already anticipated in \cite{Bork:2022vat}.

The combination of Steinmann and stampedes relations is sufficient to fix the decagon in the limit where all the neighbouring distances are becoming null. The decagon in this regime reads
\begin{equation}
\ln\mathbb{D} = \mathcal{C}_0+ \sum_{i=1}^{5}\left((-2\lambda+8\zeta_2\lambda^2)\ln{u_i}\ln{u_{i+1}}+4\zeta_2\lambda^2(\ln{u_i}\ln{u_{i+2}}+\ln^2{u_i})\right)
\label{decNull}
\end{equation}
where\footnote{This constant is the hardest to get, since it depends on the complicated part of the asymptotic regions of the integrals, the one that is given in terms of master integrals and do not generate logs.} $\mathcal{C}_0= -10\zeta_2\lambda +135\zeta_4\lambda^2$. Unfortunately, using only these two constraints is not sufficient to fix the decagon for general kinematics. At this stage, the full constraint of the decagon can only be achieved by restricting our expression to the plane kinematics and comparing with the expressions of \cite{Fleury:2020ykw}. This fixes all the polynomials $a_i,b_i,c_i$ as rational functions of cross-ratios and yields the result \eqref{decResult}, but it is not a promising strategy for fixing higher loops or higher polygons, since it requires performing complicated integrability computations.

\section{Discussion}
It has been known for more than a decade that correlation functions of the lightest scalars in planar gauge theories are dual,  when the points approach the cusps of a null polygon, to a Wilson loop evaluated in the null polygon path \cite{Alday:2010zy}. In $\mathcal{N}=4$ SYM, this relation is enhanced to an even more interesting triality since null polygonal Wilson loops are also related to gluon scattering amplitudes \cite{Alday:2007hr}. 

More recently, this triality was promoted, using conformal bootstrap methods\cite{Alday:2013cwa,Bercini:2020msp,Bercini:2021jti}, to a \textit{quadrality} with the inclusion of correlation functions of large spinning operators in the relations. A crucial point in the bootstrap derivation relies on the assumption that the correlators of spinning operators have a simple behaviour in the large spin limit. The Lorentzian inversion formula\cite{Caron-Huot:2017vep} puts this assumption on firm ground for the simplest case when only one spinning operator is involved. Unfortunately, for two or more spinning operator a generalization of the Lorentzian inversion formula is not available yet and one has to resort to specific theories to get support for these assumptions. This is precisely what we have accomplished in this paper by looking into $\mathcal{N}=4$ SYM at two loops for the next to simplest correlator, {\it i.e.} three point functions of two spinning operators and one scalar (which should be dual to a null polygon Wilson loop with five edges \cite{Bercini:2020msp}). This two loop exercise is particularly important, given the existence of a partial result in the literature \cite{Bianchi:2019jpy} that appears to be in contradiction with the usual assumptions of the analytic conformal bootstrap methods. 

We have computed this two loop three point function, resolved the apparent contradiction and showed that the assumptions of the conformal bootstrap methods hold (at least for this model and at two loop order).  These OPE coefficients involving two spinning operators were extracted by doing conformal block decomposition of a five point correlation function. Besides the large spin structure, we also obtained more than a thousand three point functions for for several finite spins. More importantly, our method can be used to obtain even more if they are needed.  

The main difficulty in carrying out the computation was that the five point correlator was expressed in terms of some conformal integrals that were previously unknown. So a by-product of our endeavor was the evaluation of two loop five point conformal integrals in particular limits, obtained using two different methods. The computation of these integrals is important per se as they appear in other physical quantities. Moreover, we have observed interesting properties of these two loop conformal integrals. For example, all two loop five point integrals can be expressed in terms of multiple polylogarithms (in the plane kinematics) or that in the open-box limit the result can be expressed in terms of harmonic polylogarithms. 

A natural and interesting next step would be to check if these properties continue to hold for higher point functions or higher loops. In fact, studying these type of conformal integrals at six points would be interesting from the WL/correlation function function duality since WL is not only given by the BDS-ansatz and starts to depend on finite cross-ratios \cite{Bern:2005iz}. Testing these relations beyond one loop would be extremely non-trivial check of the duality and could be useful to establish integrability/bootstrap understanding of this web of relations. 

Still at five points we considered null polygonal configurations of a different type of correlator, the decagon. A remarkable difference between the decagon and the light correlator in the null pentagon limit can be seen when comparing equations \eqref{nullC} and \eqref{decNull}. While the divergent part of the light correlator is governed by the usual $\Gamma_{\text{cusp}}= 8\lambda -16\zeta_2\lambda^2+\dots$ the divergent part of the decagon is governed by $\Gamma_{\text{oct}} = 8\lambda -32\zeta_2\lambda^2+\dots$, as anticipated by \cite{Bork:2022vat}. These distinct UV divergences of the correlator get translated into distinct IR divergences of the amplitudes. As explained in \cite{Caron-Huot:2021usw} this difference can be understood in terms of Coulomb branch regularizations of the amplitudes \cite{Alday:2009zm}, depending on the order that one sends the external and internal masses of the Coulomb branch amplitudes to zero.

It would be fascinating to extend this analysis of the decagon to higher loop orders, or perhaps to fully bootstrap this object as the octagon. The central problem with this approach is the lack of a simple computable basis. Using the results of our integrals we reconstructed the decagon in several kinematics, but in none of them we were able to come up with a simple basis of integrals or functions (akin to the ladders of the octagon) that we could kick start the bootstrap games. In fact, this is not surprising given that even the octagon at two loops would also not give a strong hint of what the all loop basis could be. Higher loops results played a much more decisive role in hinting what the all loop basis turned out to be. 

The only alternative at the moment is to continue to work with the basis of all possible five point conformal integrals. If one can compute the log divergence of these integrals then, in principle, one could play the same bootstrap games as we did. First use Steinmann relations to reduce the number of integrals and then stampedes to fix the coefficients of these integrals, ending up with an expression for the null limit of the decagon. At two loops this is enough to fix all the coefficients, but there is no guarantee that this is sufficient to constrain the coefficients at higher loops.

Both Steinmann and stampedes can be easily extended to include more points, therefore this bootstrap games could be also extended to higher point functions. For instance, for six point functions one could start to probe even more complicated objects such as the \textit{dodecagon}. Where many intriguing aspects such as connections to amplitudes and origins limits could be studied \cite{Basso:2020xts}. We are starting to explore the complicated world of higher point functions at higher orders in perturbation theory, and we are intrigued to see if our one loop intuition remains correct.

\section*{Acknowledgements} 
We would like to thank Till Bargheer, Frank Coronado, Albert Bekov and Antonio Antunes for illuminating
discussions. Centro de F\'{i}sica do Porto is partially funded by Funda\c{c}\~{a}o para a Ci\^{e}ncia e a Tecnologia (FCT) under the grant UID04650-FCUP. The work of CB was funded by the Deutsche Forschungsgemeinschaft (DFG, German Research Foundation) -- 460391856. V.G. is supported by Simons Foundation grants \#488637 (Simons collaboration on the non-perturbative bootstrap) and Fundacao para a Ciencia e Tecnologia (FCT) under the grant CEECIND/03356/2022. B.F. is supported by Simons Foundation grant \#488637 (Simons collaboration on the non-perturbative bootstrap) and by Funda\c{c}\~{a}o para a Ci\^{e}ncia e Tecnologia, under the IDPASC doctoral program, under the grand PRT/BD/154692/2022.
\appendix 
\section{Master integral and form of the differential equation}\label{app:masterintegrals}
Here we list the 28 master integrals $g_i$ appearing in equation \eqref{eq:DintHardRegion}. The differential equation that they satisfy has this form
\begin{align}
\frac{d}{d u_j} g_i = \sum_{m}A^{j}_{im}g_{m}
\end{align}
where the matrices $A^{j}_{im}$ are rational functions of the cross ratios and $\epsilon$ and are given in an auxiliary file. The master integrals $g_i$ are defined by

\begin{align*}
	&g_1 = \int \frac{d^dx_6d^dx_7}{x_{17}^2 x_{18}^2 x_{37}^2x_{38}^2}, &g_2=&\int \frac{d^dx_6d^dx_7}{x_{17}^2 x_{28}^2 x_{37}^2 x_{38}^2}, &g_3=&\int\frac{d^dx_6d^dx_7}{x_{17}^2 x_{18}^2 x_{38}^2 x_{47}^2},\\
	&g_4 = \int \frac{d^dx_6d^dx_7}{x_{28}^2 x_{37}^2 x_{78}^2}, &g_5=&\int \frac{d^dx_6d^dx_7}{x_{18}^2 x_{47}^2 x_{78}^2}, &g_6 =& \int \frac{d^dx_6d^dx_7}{x_{18}^2 x_{38}^2 x_{47}^2 x_{78}^2}, \\ &g_7=\int\frac{d^dx_6d^dx_7}{x_{17}^2 x_{28}^2 x_{37}^2 x_{78}^2}, &g_8 =&\int\frac{d^dx_6d^dx_7}{x_{17}^2 x_{28}^2 x_{37}^2 x_{38}^2 x_{78}^2}, &g_9=&\int \frac{d^dx_6d^dx_7}{x_{17}^2 x_{18}^2 x_{38}^2 x_{47}^2 x_{78}^2}, \\
	&g_{10} = \int \frac{d^dx_6d^dx_7}{x_{28}^2 x_{38}^2 x_{47}^2 x_{78}^2},  &g_{11} =&\int \frac{d^dx_6d^dx_7}{x_{18}^2 x_{37}^2 x_{78}^2},  &g_{12}=&\int \frac{d^dx_6d^dx_7}{x_{28}^2 x_{47}^2 x_{78}^2},\\
	&g_{13} = \int \frac{d^dx_6d^dx_7}{x_{17}^2 x_{28}^2 x_{38}^2 x_{78}^2}, &g_{14}=&\int\frac{d^dx_6d^dx_7}{x_{17}^2 x_{28}^2 x_{38}^2 x_{47}^2},  &g_{15}=& \int \frac{d^dx_6d^dx_7}{x_{1,7}^2 x_{3,8}^2 x_{4,7}^2 x_{7,8}^2}, \\
	&g_{16} = \int \frac{d^dx_6d^dx_7}{x_{17}^2 x_{28}^2 x_{47}^2 x_{78}^2},  &g_{17} =& \int \frac{d^dx_6d^dx_7}{x_{17}^2 x_{28}^2 x_{38}^2 x_{47}^2 (x_{78}^2)^2},  &g_{18}=&\int \frac{d^dx_6d^dx_7}{x_{17}^2 x_{28}^2 x_{38}^2 (x_{47}^2)^2 x_{78}^2}, \\
	&g_{19} = \int \frac{d^dx_6d^dx_7}{x_{17}^2 (x_{28}^2)^2 x_{38}^2 x_{47}^2 x_{78}^2}, &g_{20}=& \int \frac{d^dx_6d^dx_7}{(x_{17}^2)^2 x_{28}^2 x_{38}^2 x_{47}^2 x_{78}^2}, &g_{21}=& \int \frac{d^dx_6d^dx_7}{x_{17}^2 x_{28}^2 x_{38}^2 x_{47}^2 x_{78}^2},\\
	&g_{22}=\int \frac{d^dx_6d^dx_7}{x_{17}^2 x_{18}^2 x_{28}^2 x_{38}^2 x_{47}^2 x_{78}^2}, &g_{23}=& \int \frac{d^dx_6d^dx_7}{x_{18}^2 x_{28}^2 x_{37}^2 x_{47}^2 x_{78}^2}, &g_{24}=&\int \frac{d^dx_6d^dx_7}{x_{17}^2 x_{28}^2 x_{37}^2 x_{38}^2 x_{47}^2 x_{78}^2},\\
	&g_{25}=\int \frac{d^dx_6d^dx_7}{x_{17}^2 x_{28}^2 x_{37}^2 x_{47}^2 x_{78}^2}, &g_{26}=&\int \frac{d^dx_6d^dx_7}{x_{18}^2 x_{28}^2 x_{38}^2 x_{47}^2 x_{78}^2}, &g_{27}=& \int \frac{d^dx_6d^dx_7}{x_{17}^2 x_{18}^2 x_{28}^2 x_{37}^2 x_{38}^2 x_{47}^2 x_{78}^2},\\
	&g_{28}= \int \frac{d^dx_6d^dx_7}{x_{17}^2 x_{18}^2 x_{28}^2 (x_{37}^2)^2 x_{38}^2 x_{47}^2 x_{78}^2}.
\end{align*}
One important one loop master integral is given by
\begin{align}
&\int \frac{d^dx_8}{x_{18}^2 x_{38}^2 x_{48}^2}=   \frac{12 \text{Li}_2(1-v)+6 \ln u \ln v}{6 (1-v)}+\epsilon\bigg[\frac{2 \text{Li}_3\left(\frac{v-1}{v}\right) -2 \text{Li}_3(1-v)}{(v-1)} \label{f0eq} \\
&+\frac{12 \text{Li}_2(1-v) (\ln u+2)+3 \ln u \ln ^2v+3 \ln ^2u \ln v+12 \ln u \ln v-2 \ln ^3v}{6 (v-1)}\bigg]+O(u)\nonumber
\end{align}
where we have taken  $x_{13}^2=1$ and used $u=x_{34}^2,v=x_{14}^2$. Note that we have only written the leading order in $u\rightarrow 0$ limit since we only needed this order. 
\section{Perturbative data}
\label{appData}
Here we review what is known about the perturbative expressions of structure constants of one and two spin operators up to two loops, as well as establish conventions used in throughout the paper.

The structure constants for one spinning operator have been studied extensively in the literature. The most efficient way of computing these three point functions are via the conformal block decomposition of a four point function. The first non trivial computation of these three point functions in $\mathcal{N} = 4$ was done at one loop level in \cite{Dolan:2004iy} and later extended for up to three loops and any spin in \cite{Eden:2012rr} and for spin two operators is known up to five loops \cite{Georgoudis:2017meq}. Here we recall their expressions up to two loops
\begin{align}
C^{\bullet\circ\circ}_J &= C^{\bullet\circ\circ}_\texttt{tree}(1+\lambda\hat{C}^{\bullet\circ\circ}_\texttt{1-loop}+\lambda^2\hat{C}^{\bullet\circ\circ}_\texttt{2-loop}+\dots)\\
C^{\bullet\circ\circ}_\texttt{tree} &= \frac{J!}{\sqrt{(2J)!}} \\
C^{\bullet\circ\circ}_\texttt{1-loop} &= 4S_1(J)^2-4S_1(J)S_1(2J)-2S_2(J)\\
C^{\bullet\circ\circ}_\texttt{2-loop} &= 20 S_{-4}(J)+8S_{-2}(J)^2+8S_{-3}(J)S_{1}(J)-16S_{-2}(J)S_{1}(J)^2+8S_{1}(J)^4+\nonumber\\
&+8S_{-3}(J)S_{1}(2J)+16S_{-2}(J)S_{1}(J)S_{1}(2J)-16S_{1}(J)^3S_{1}(2J)+8S_{1}(J)^2S_{1}(2J)^2+\nonumber\\
&+8S_{-2}(J)S_{2}(J) -32S_{1}(J)^2S_{2}(J)+24S_{1}(J)S_{1}(2J)S_{2}(J)+16S_{1}(J)^2S_{2}(2J)+\nonumber\\
&+6S_{2}(J)^2+8S_{1}(J)S_{3}(J) +8S_{1}(2J)S_{3}(J)+20S_{4}(J)+16S_{1}(J)S_{-2,1}(J)+\nonumber\\
&-16S_{-3,1}(J)+8S_{-2,2}(J)-16S_{1,3}(J)+24\zeta_3S_{1}(J)
\end{align}
where the $S_{i_1,\dots,i_n}(J)$ are the harmonic sums.

To set the coupling convention we also recall the anomalous dimension of these operators
\begin{align}
\gamma(J) &= \lambda \gamma_{\texttt{1-loop}}+\lambda^2\gamma_{\texttt{2-loop}} + \dots \\
\gamma_{\texttt{1-loop}} &= 8S_{1}(J) \\
\gamma_{\texttt{2-loop}} &=-16S_{-3}(J)-32S_{-2}(J)S_{1}(J)-32S_{1}(J)S_{2}(J)-16S_{3}(J)+32S_{-2,1}(J)
\end{align}
so at large spin we have the following cusp and collinear anomalous dimension
\begin{align}
f(\lambda) &=8\lambda -16\zeta_2\lambda^2+\dots \\
g(\lambda) &=8\gamma_E\lambda -(24\zeta_3-16\zeta_2\gamma_E)\lambda^2+\dots
\end{align}

For spinning operators we will use the usual index free notation, which consist in contracting the spinning operator with an auxiliary null polarization vector $\epsilon$ 
\begin{equation}
\mathcal{O}_J(x,\epsilon) = \epsilon_{\mu_1}\dots\epsilon_{\mu_J}\mathcal{O}^{\mu_1\dots\mu_J}(x)
\end{equation}
The three point function of two such operators is parametrized as
\begin{equation}
\langle \mathcal{O}_{J_1} (x_1, \epsilon_1), \mathcal{O}_{J_2} (x_2, \epsilon_2) , \mathcal{O}(x_3) \rangle =
\frac{V_{1,23}^{J_1} V_{2,31}^{J_2}}{x_{12}^{\kappa_1 + \kappa_2 - \kappa_3}x_{23}^{\kappa_2 + \kappa_3 - \kappa_1}x_{31}^{\kappa_3 + \kappa_1 -\kappa_2}}\sum_{\ell=0}^{\ell_{\text{max}}}C^{\bullet\bullet\circ}_{J_1,J_2,\ell}T^\ell
\label{TwoSpinning3pt}
\end{equation}
where  $\kappa_i = \Delta_i+J_i$ is the conformal spin and
\begin{equation} V_{i,jk} =\frac{\epsilon_i \cdot  x_{ik} x_{ij}^2 - \epsilon_i \cdot  x_{ij} x_{ik}^2}{x^2_{jk}},  \quad  H_{ij} = (\epsilon_i \cdot x_{ij})( \epsilon_j \cdot x_{ij}) - \tfrac{1}{2} x^2_{ij}(\epsilon_i \cdot \epsilon_j)\,,\quad T = \frac{H_{12}}{V_{1,23}V_{2,31}}
\end{equation}
are a basis of conformal covariant tensors defined in \cite{Costa:2011mg}. The three point function is a polynomial in this tensor structures with the degree fixed by conformal symmetry to be $\ell_{\text{max}}=\text{min}(J_1,J_2)$.

The structure constants of two spinning operators appearing in this three point function were also worked out up to one loop \cite{Bianchi:2019jpy,Bercini:2021jti}.
\begin{align*}
C^{\bullet\bullet\circ}_{J_1,J_2,\ell} &= C^{\bullet\bullet\circ}_\texttt{tree}(1+\lambda\hat{C}^{\bullet\bullet\circ}_\texttt{1-loop}+\lambda^2\hat{C}^{\bullet\bullet\circ}_\texttt{2-loop}+\dots)\\
C^{\bullet\bullet\circ}_\texttt{tree} &= \frac{J_1!}{\sqrt{(2J_1)!}}\frac{J_2!}{\sqrt{(2J_2)!}}\binom{J_1}{\ell}\binom{J_2}{\ell} \\
C^{\bullet\bullet\circ}_\texttt{1-loop} &= 4\left(S_1(J_1)+S_1(J_2)\right)\sum_{i=1}^{\ell}\frac{\binom{J_1+J_2+1}{i}\binom{\ell}{i}}{i \binom{J_1}{i}\binom{J_2}{i}}-4 \frac{(J_1+1)(J_2+1)}{J_1+J_2+2}\sum_{i=1}^\ell\frac{\binom{J_1+J_2+2}{i}a_i^\ell}{i \binom{J_1}{i}\binom{J_2}{i}}+ \nonumber \\
& +4S_1(J_1)^2+-4S_1(J_1)S_1(2J_1)-2S_2(J_1)+4S_1(J_2)^2-4S_1(J_2)S_1(2J_2)
-2S_2(J_2)\\
C^{\bullet\bullet\circ}_\texttt{2-loop} &= \dots + 24\zeta_3|S_1(J_1)-S_1(J_2)|
\end{align*}
where the constants $a_i^\ell$ are given by
\begin{equation}
a_i^\ell= \frac{(-1)^i}{2} \left( -S_1(\ell)^2 - S_2(\ell) +
2\sum_{k=2}^{i} \frac{(-1)^k\binom{l}{k-1}}{k-1}\left(S_1(k-2)+S_1(\ell)\right)\right) \,.
\label{as}
\end{equation}
and the two loop structure constants are unknown up to the $\zeta_3$ coefficient.

In \cite{Bercini:2022gvs} a new basis for the three point functions was introduced, it reads
\begin{equation}
\langle \mathcal{O}_{J_1} (x_1, \epsilon_1), \mathcal{O}_{J_2} (x_2, \epsilon_2) , \mathcal{O}(x_3) \rangle =
\frac{V_{1,23}^{J_1} V_{2,31}^{J_2}}{x_{12}^{\kappa_1 + \kappa_2 - \kappa_3}x_{23}^{\kappa_2 + \kappa_3 - \kappa_1}x_{31}^{\kappa_3 + \kappa_1 -\kappa_2}}\sum_{\ell=0}^{\ell_{\text{max}}}A^{\bullet\bullet\circ}_{J_1,J_2,\ell} T^\ell (1-T)^{\ell_{\text{max}}-\ell}
\label{TwoSpinning3ptNew}
\end{equation}
we denominate it as the integrability basis, since it makes trivial the relation of the integrable hexagons \cite{Basso:2015zoa} with the spinning structure constants, see \cite{Bercini:2022gvs}. The map between the usual basis and integrability basis is also vey simple to work out.
\begin{equation}
A(J_1,J_2,\ell) =\sum_{k=0}^\ell \frac{\binom{\ell_{\text{max}}}{\ell}\binom{\ell}{k}}{\binom{\ell_{\text{max}}}{k}}C(J_1,J_2,k)
\end{equation}
\begin{equation}
C(J_1,J_2,\ell) =\sum_{k=0}^\ell \frac{\binom{\ell_{\text{max}}}{\ell}\binom{\ell}{k}}{\binom{\ell_{\text{max}}}{k}}(-1)^{\ell-k}A(J_1,J_2,k)
\label{CvsA}
\end{equation}

One can also work out the two spin structure constants in teh integrability basis
\begin{align*}
A^{\bullet\bullet\circ}_{J_1,J_2,\ell} &= A^{\bullet\bullet\circ}_\texttt{tree}(1+\lambda\hat{A}^{\bullet\bullet\circ}_\texttt{1-loop}+\lambda^2\hat{A}^{\bullet\bullet\circ}_\texttt{2-loop}+\dots)\\
A^{\bullet\bullet\circ}_\texttt{tree} &= \frac{J_1!}{\sqrt{(2J_1)!}}\frac{J_2!}{\sqrt{(2J_2)!}}\binom{J_{\text{min}}}{\ell}\binom{J_{\text{max}}+\ell}{\ell} \\
A^{\bullet\bullet\circ}_\texttt{1-loop} &=6S_1({J_{\text{min}}})^2-4S_1({J_{\text{min}}})S_1(2{J_{\text{min}}})
-4S_1({J_{\text{max}}})S_1(2{J_{\text{max}}})
-4S_1({J_{\text{max}}})S_1({J_{\text{min}}}-{\ell})+\\
&-2S_1({J_{\text{min}}}-{\ell})^2-4S_1({J_{\text{min}}})S_1({\ell}) +4S_1({J_{\text{min}}}-{\ell})S_1({\ell})-2S_1({\ell})^2-2S_2({\ell})+\nonumber\\
&+4S_1({J_{\text{min}}})S_1({J_{\text{max}}}+{\ell})+4S_1({J_{\text{max}}})S_1({J_{\text{max}}}+{\ell})-2S_2({J_{\text{max}}})
-2S_2({J_{\text{min}}}-{\ell}) \\ 
A^{\bullet\bullet\circ}_\texttt{2-loop} &= \dots + 24\zeta_3\binom{J_\text{min}+1}{\ell}(S_1(J_\text{max})-S_1(J_\text{min}))
\end{align*}

It is amazing to see that this simple combination of binomials can transform the ugly sums of the usual basis into an expression that has only nice harmonic numbers of the new basis. It also has similar features as \cite{Costa:2023wfz}, since polarizations and spins are in the same footing one can expect analyticity of the structure constants in these quantities in large spin and polarizations limits like the Regge limit studied in \cite{Costa:2023wfz}.

\section{Examples of conformal integrals}
\label{appInts}
We start by recalling the definition of the multiple polylogarithms \cite{Goncharov:1998kja} via the recursive iterated integral 
\begin{equation}
G(a_1,\dots,a_n;z) = \int_{0}^{z} \frac{dt}{t-a_1}G(a_2,\dots,a_n;t)
\end{equation}
with $G(z)=1$ and $a_i,z$ are complex variables.


We also consider four point integrals at one and two loops, they are defined as
\begin{align}
\Phi^{(1)}(u_{\textcolor{c1}{1}} u_{\textcolor{c4}{4}},u_{\textcolor{c5}{5}}) &= x_{\textcolor{c1}{1}\textcolor{c4}{4}}^2x_{\textcolor{c2}{2}\textcolor{c5}{5}}^2\int \frac{d^{4}x_{6}}{x_{\textcolor{c1}{1}6}^2x_{\textcolor{c2}{2}6}^2x_{\textcolor{c4}{4}6}^2x_{\textcolor{c5}{5}6}^2} \\
\Phi^{(2)}(u_{\textcolor{c1}{1}} u_{\textcolor{c4}{4}},u_{\textcolor{c5}{5}}) &= x_{\textcolor{c1}{1}\textcolor{c4}{4}}^4x_{\textcolor{c2}{2}\textcolor{c5}{5}}^2\int \frac{d^{4}x_{6}d^{4}x_{7}}{x_{\textcolor{c1}{1}6}^2x_{\textcolor{c2}{2}6}^2 x_{\textcolor{c5}{5}6}^2  x_{\textcolor{c2}{2}7}^2 x_{\textcolor{c4}{4}7}^2 x_{\textcolor{c5}{5}7}^2 x_{67}^2}
\end{align}
and can be easily integrated
\begin{align}
    \Phi^{(L)}(u,v) &= \frac{1}{\bar{z}-z}F^{(L)}\left(\frac{z}{1-z},\frac{\bar{z}}{1-\bar{z}}\right)\,,\\
    F^{(L)}(z,\bar{z}) &= \sum_{k=0}^{L}\frac{(-1)^{k}(2L-k)!}{k!(L-k)!L!}\ln^{k}(z\bar{z})\left(\text{Li}_{2L-k}(z)-\text{Li}_{2L-r}(\bar{z})\right)
\end{align}
being $z,\bar{z}$ are the usual four point cross-ratios $z\bar{z} = u$ and $(1-z)(1-\bar{z})=v$.

In general we cannot write in such small expressions the five point two loop integrals. However in the kinematical regimes described in the figure \ref{figChannels} we can. For example, in the open-box limit depicted on the left of figure \ref{figChannels} one of the integral reads
\begin{equation}
\mathbb{L}_{2,13,45} =\frac{7\pi^4}{180} +\frac{\pi^2}{2}\ln^2{u_1}+\frac{\pi^2}{6}\ln{u_1}\ln{u_4} -3 \zeta_3\ln{u_1} -3\zeta_3\ln{u_4} +\dots
\end{equation}
where the dots represent subleading contributions. In the OPE limit depicted on the right of figure \ref{figChannels} another example of integral is given by
\begin{align}
\mathbb{D}_{13,245} &= 4+U_4-U_5+\frac{11}{6}U_4U_5 -\zeta_3U_4U_5 + \ln{u_1}\left(-2+\frac{3}{2}U_4-\frac{U_5}{2}+\frac{U_4U_5}{2}\right) + \nonumber\\
&+\ln{u_3}\left(-2+\frac{U_4}{2}-\frac{U_5}{2}+\frac{U_4U_5}{2}\right)+\ln{u_1}\ln{u_3}\left(1+\frac{U_4}{2}+\frac{U_5}{2}-\frac{U_4U_5}{4}\right)+\dots
\end{align}
where $\mathbb{D}_{13,245} = x^2_{24}x^2_{25}x^2_{45}\mathcal{D}_{13,245}$ is defined using \eqref{intD} to be a function only of cross-ratios and $U_i=(1-u_i)$ are the cross-ratios that are expanded around zero in the OPE expansion.

\bibliography{references}
\bibliographystyle{nb}

\end{document}